\documentclass[a4paper,11pt]{article}

\usepackage[utf8x]{inputenc}
\usepackage[T1]{fontenc}
\usepackage{color}
\usepackage{amssymb}
\usepackage{amsmath}
\usepackage{graphicx}
\usepackage{mathtools}
\usepackage{ragged2e}
\DeclareMathOperator{\sech}{sech}

\usepackage{dcolumn}
\usepackage{bm}
\usepackage{graphicx}
\usepackage{braket}
\usepackage{verbatim} 
\usepackage[english]{babel}
\usepackage{hyperref}
\hypersetup{
citecolor=red,
colorlinks=true,
filecolor=red,
linkcolor=blue,
linktocpage=true,
urlcolor=blue
} 
\usepackage[titletoc,toc,title]{appendix}
\numberwithin{equation}{section}
\usepackage{cleveref}
\usepackage[margin=2.5cm]{geometry}
\usepackage{cite}
\usepackage{epsfig}
\usepackage{float}
\usepackage{amsfonts}
\usepackage{enumitem}
\usepackage[font={footnotesize,it}]{caption}

\numberwithin{equation}{section}

\begin{document}

\begin{flushright}
APCTP Pre2023 - 013
\end{flushright}

\begin{center}

\centerline{\Large {\bf Effect of backreaction on Island, Page curve and Mutual Information}}

\vspace{8mm}

\renewcommand\thefootnote{\mbox{$\fnsymbol{footnote}$}}
Parul Jain,${}^{1}$\footnote{parul24jain@gmail.com} 
Sanjay Pant,${}^{2,3}$\footnote{sanjaypant.phy@geu.ac.in} and Himanshu Parihar ${}^{4,5}$
\footnote{himansp@phys.ncts.ntu.edu.tw} 

\vspace{4mm}

${}^1${\small \sl Asia Pacific Center for Theoretical Physics} \\
{\small \sl Pohang 37673, Republic of Korea} 
\vskip 0.2cm

${}^2${\small \sl Department of Physics}\\
{\small \sl Indian Institute of Technology Ropar}\\
{\small \sl  Rupnagar, Punjab 140001, India} 

\vskip 0.2cm
${}^3${\small \sl Department of Allied Sciences (Physics)}\\
{\small \sl Graphic Era (Deemed to be University)}\\
{\small \sl  Dehradun, Uttarakhand 248002, India} 
\vskip 0.2cm

${}^4${\small \sl Center of Theory and Computation}\\
{\small \sl National Tsing-Hua University}\\
 {\small \sl Hsinchu 30013, Taiwan} 
 \vskip 0.2cm
${}^5${\small \sl Physics Division}\\
   {\small \sl  National Center for Theoretical Sciences}\\
     {\small \sl Taipei 10617, Taiwan} 
\vskip 0.2cm

\end{center}

\vspace{6mm}
\numberwithin{equation}{section}
\setcounter{footnote}{0}
\renewcommand\thefootnote{\mbox{\arabic{footnote}}}

\begin{abstract} 
{ We compute the entanglement entropy of Hawking radiation in a bath attached to a deformed eternal AdS black hole. This black hole is dual to the two identical strongly coupled large-$N_c$ thermal field theories, where each theory is backreacted (deformed) by the presence of a uniform static distribution of heavy fundamental quarks. In our observation we find that the entanglement entropy of Hawking radiation increases in a quadratic manner for an early time and linearly for the late time. The large time expression for the entanglement entropy of Hawking radiation is used to find the Page curve and Page time. After the Page time, the entanglement entropy saturates to a constant value due to the appearance of an island. We observe that introducing deformation (backreaction) delays the appearance of island and shifts the Page curve to a later time. Subsequently, the computation of the scrambling time reveals an increase with the backreaction parameter, suggesting a longer duration for information retrieval in the presence of deformation. Moreover, our analysis of the mutual information between the radiation subsystems shows that it vanishes at a critical time which increases with the deformation before the Page time. After the Page time, the appearance of island leads to the vanishing of mutual information between black hole subsystems and gives the time difference of the order of scrambling time.}

\end{abstract}

\newpage
\tableofcontents
\newpage

\section{Introduction}\label{sec:intro}

The black hole information loss paradox is one of the most important and fascinating problem that has profound implications for our understanding of the quantum gravity \cite{Hawking:1975vcx,Hawking:1976ra}. Given that black hole radiation resembles thermal radiation, it has been demonstrated that during the evaporation of a black hole, the entropy of the radiation increases monotonically \cite{Hawking:1975vcx}.
The expectation from unitarity is that the radiation's entropy should start decreasing after reaching a certain value and eventually becomes zero when the black hole completely evaporates. This expected behavior stems from the fact that the quantum fields are supposed to be in a pure state before and after the black hole evaporation process.
The variation of the radiation's entropy with time which is depicted by the Page curve introduces a characteristic timescale known as the Page time \cite{Page:1993wv,Page:2013dx}. Focusing on the Page curve, the information loss paradox can be seen in the following heuristic way. Consider a Hawking pair created near the black hole horizon with one particle falling in and the other escaping to a spacetime region called the radiation bath. The von Neumann entropy of the radiation is initially zero due to no Hawking pairs and increases as the black hole evaporates due to the accumulation of more particles in the bath. The entropy of the emitted radiation increases consistently in a monotonic manner, and at some point, it becomes greater than the Bekenstein-Hawking entropy \cite{Hawking:1975vcx, Hawking:1976ra}. Since the Bekenstein-Hawking entropy is the maximum amount of entropy that a black hole can possess, this behavior leads to the information paradox. In other words, the fine grained entropy of the radiation is given by the von Neumann entropy of the quantum fields existing outside the black hole. Now taking into account the fact that the quantum fields are in a pure state, this fine grained entropy is expected to be less than or equal to the coarse grained entropy (Bekenstein-Hawking entropy) of the black hole. This paradox happens during the black hole evaporation after the Page time when the fine grained entanglement entropy of the radiation dominates over the coarse grained entropy of the black hole implying unitarity violation. According to the suggestion proposed by D. Page, if the evaporation process is unitary, one can expect that the entropy of radiation increases until the von Neumann entropy becomes equal to the Bekenstein-Hawking entropy, and after that, it should start decreasing \cite{Page:1993wv}. In particular, the Page curve shows that initially, the fine-grained entropy increases with time following the Hawking curve up to halfway during the evaporation; after that, it starts decreasing and finally vanishes.

In recent years, a lot of progress has been made to resolve the information paradox via the island prescription which involves contribution towards the fine grained entropy of the radiation from specific spacetime regions in the black hole geometry known as the ``islands'' \cite{Penington:2019npb, Almheiri:2019psf, Almheiri:2019hni, Almheiri:2019yqk, Almheiri:2019psy, Almheiri:2020cfm}. The boundaries of these islands are known as the quantum extremal surfaces (QES) \cite{Engelhardt:2014gca}. The QES comes into play when one incorporates the quantum corrections from the bulk towards the Ryu-Takayanagi (RT) formula \cite{Ryu:2006bv}. 
As unitarity is the key ingredient when dealing with black hole information paradox and in order to preserve it, radiation's entanglement entropy is expected to decrease after the Page time during the evaporation of black hole. 
In the case of an eternal black hole the black hole does not evaporate completely hence it is expected that instead of decreasing, the radiation’s entropy comes to a constant value after the Page time. 
In order to evaluate the Page curve using the island prescription, the black hole is coupled to a bath which collects the radiation. In the case of an eternal black hole, two baths are attached to the left and the right boundaries of eternal black hole in order to assess the Page curve as shown in \cref{set-up}. In the island formulation as we extremize the generalized entropy for black hole radiation which is given by the sum of the area of the boundary of island and the contribution from entanglement entropy of the quantum matter fields. This generalized entropy incorporates both the area term and bulk quantum corrections, the expression for the fine grained entropy of the Hawking radiation in a region $R$ gets modified as follows
\begin{equation}\label{island}
S(R)=\textrm{min}\bigg\{{\textrm{ext}}_{\mathcal{\mathrm{I}}}\biggl(\frac{\textrm{Area}(\partial I)}{4G_N}+S_{\textrm{matter}}(R\cup I)\biggr)\bigg\},
\end{equation}
wherein $R$ denotes the radiation, $I$ denotes the island
and $\partial I$ denotes the boundary of the island, in other words the quantum extremal surface (QES). The expression inside the bracket in \cref{island} is the generalized entropy defined on $R\cup I$. Initially, in the absence of island, the entanglement entropy of radiation keeps increasing but afterwards, at a later time, the presence of island saturates the entanglement entropy of the radiation. This modification in turns leads to the expected Page curve \cite{Page:1993wv, Page:1993df, Page:2013dx} thereby preserving unitarity. 
The island formula inspired by the quantum extremal surfaces and derived from replica wormhole saddle points in the gravitational path integral \cite{Penington:2019kki,Almheiri:2019qdq} has been extensively explored in the context of two dimensional Jackiw-Teitelboim (JT) gravity due to their relative simplicity compared to higher dimensional black holes.  Importantly, this prescription has also been validated in higher dimensional black holes \cite{Almheiri:2019psy, Hashimoto:2020cas, Wang:2021woy, Yu:2021cgi, Ahn:2021chg, Karananas:2020fwx, Lu:2021gmv, Arefeva:2021kfx, He:2021mst}. These studies also involve entanglement measures beyond the von Neumann entropy \cite{Chandrasekaran:2020qtn, KumarBasak:2020ams, KumarBasak:2021rrx, Bhattacharya:2021jrn, Li:2021dmf, Bhattacharya:2021dnd, Ling:2021vxe, Saha:2021ohr, Akers:2022max, BasakKumar:2022stg, Lin:2022qfn,Basu:2022reu, Shao:2022gpg, Afrasiar:2022ebi,RoyChowdhury:2022awr,  Afrasiar:2022fid, Afrasiar:2023jrj, RoyChowdhury:2023eol, Kumari:2023ops}. Apart from the widely investigated JT gravity setup, this development led to the analysis of the island formula and Page curve in various other scenarios \cite{Goto:2020wnk, Colin-Ellerin:2020mva, Kawabata:2021vyo,Geng:2020qvw,Geng:2020fxl,Deng:2020ent,Krishnan:2020oun, Krishnan:2020fer,Hollowood:2020cou, Chen:2020uac, Chen:2020hmv, Hernandez:2020nem,Akal:2020twv, Uhlemann:2021nhu, Yu:2021rfg, Krishnan:2021faa, Chu:2021gdb,Ghosh:2021axl, Krishnan:2021ffb, Hollowood:2021nlo, Hollowood:2021wkw,Akal:2021dqt, Omidi:2021opl, Cadoni:2021ypx, Hollowood:2021lsw, Geng:2021hlu, Yadav:2022fmo, Anand:2022mla, Gyongyosi:2022vaf,  Grimaldi:2022suv,Du:2022vvg, Yu:2022xlh, Ageev:2022qxv, HosseiniMansoori:2022hok, Yadav:2022jib, Lu:2022tmt, Lu:2022cgq, Basu:2022crn,Suzuki:2022xwv, Karch:2022rvr, Cadoni:2023tse, Piao:2023vgm, Guo:2023gfa, Parvizi:2023foz, Hung:2023mbw, Jeong:2023hrb, Tong:2023nvi, Yu:2023whl, Matsuo:2023cmb, Yadav:2022mnv, Matsuo:2020ypv, Iizuka:2021tut, Anegawa:2020ezn, Almheiri:2021jwq,Afrasiar:2023nir, Anand:2023ozw, Kashyap:2023keo, Blommaert:2023vbz, Guo:2023fly, Chang:2023gkt, Li:2023rue, Li:2023fly, Akers:2022qdl, Murdia:2022giv, Chandra:2022fwi, Li:2023nfv, Hirano:2023ebw,Hirano:2021rzg,Okuyama:2021bqg, Chou:2021boq, Chou:2023adi, Geng:2022slq,Gyongyosi:2023sue, Basu:2023wmv}.

On a different note, there has been a lot of interest in the study of strongly coupled gauge theory using the probe approximation. Exploring the boundary gauge theory with a substantial number of flavor quarks, especially in the context of strong coupling, continues to pose significant challenges. According to the AdS/CFT correspondence, incorporating flavor quarks into the boundary theory is equivalent to introducing an additional stack of $N_f$ flavor branes that probe the pre-existing $N_c$ color branes in the corresponding dual gravity \cite{Karch:2002sh}. In this context, the authors \cite{Chakrabortty:2011sp, Chakrabortty:2016xcb} considered a $d$ dimensional strongly coupled large $N_c$ gauge theory at finite temperature in presence of  uniform distribution of large number $N_f$ of externally added heavy fundamental quarks. When $N_f$ is approximately equal to or greater than $N_c^2$, the external heavy quark induces a notable backreaction. We refer this system as the quark cloud model. The holographic dual of the gauge theory influenced by the presence of heavy stationary fundamental quarks (quark cloud model) is an AdS black hole where strings extend from the boundary to the horizon. The end points of the strings are realised as the quarks, and the density of the quark cloud is holographically equivalent to the density of the string cloud. The distribution of strings within the string cloud is uniform, and any interactions between the strings are disregarded. The coexistence of a string cloud and a AdS black hole leads to the emergence of a deformed AdS black hole geometry in $d + 1$ dimensions as suggested in the work of \cite{Chakrabortty:2011sp}. The author in \cite{Chakrabortty:2011sp} explored the thermodynamic stability and demonstrated that the density of the string cloud is positive. In this direction, the effect of backreaction on the dissipative force on an external quark generated by the heavy stationary quark cloud was presented in \cite{Chakrabortty:2011sp}. Furthermore, in \cite{Chakrabortty:2016xcb}, authors delve into the effects of backreaction on the hydrodynamical properties of a thermal plasma within the context of $N = 4$ strongly coupled super Yang-Mills (SYM) theory. 
Interestingly, this gravitational dual of the quark cloud model has also been used as a setup to study the effects of deformation (backreaction) on the various entanglement measures as this backreacted geometry induces significant and interesting corrections. The comprehensive analysis of how backreaction influences various entanglement measures like entanglement entropy and entanglement of purification has been provided in \cite{Chakrabortty:2020ptb}.
Subsequently in \cite{Chakrabortty:2022kvq}, the authors considered two boundary field theories with each theory backreacted by the presence of a uniform distribution of heavy static fundamental quarks in a TFD state termed as the backreacted TFD state.  The corresponding bulk dual theory consists of two entangled AdS black holes, each of which is deformed by a uniform distribution of long fundamental strings attached to the boundary. 
They analysed the the impact of backreaction on the two-sided Mutual Information often referred to as thermo mutual information (TMI) in this deformed eternal black hole geometry and found that the backreaction leads to an increase of chaos in the system. Taken together, the existing literature strongly suggests that the backreaction brings significant changes in the structure of entanglement within the system.

The above developments lead naturally to inquire about the implications of backreaction on the construction of the entanglement entropy island and its consequences on the Page curve. In this article, we address this interesting issue by considering the holographic dual of a backreacted TFD state which is described by a deformed eternal (two sided) black hole glued with two auxiliary thermal baths on either side. Here each side of the black hole is deformed by a uniform distribution of static strings. Note that attaching the non-gravitational bath systems breaks the diffeomorphism symmetry leading to a massive graviton as described in \cite{Geng:2020qvw}. In this scenario, taking the zero-mass limit for the graviton result in the disappearance of any island contributions. Subsequently the authors in \cite{Geng:2021hlu} argue that the entanglement islands can exist only in massive gravity and it does not constitute consistent entanglement wedges in massless gravity obeying the gravitational Gauss law.
While the applicability of the QES formula in our setup remains an open question, we proceed under the assumption that it holds true for asymptotically flat spacetime with a massive graviton. We compute the entanglement entropy of the Hawking radiation in a bath coming from the deformed eternal black hole using the island formula. We then study the effect of backreaction parameter on the behavior of the Page curve and its impact on the emergence of the island. It is observed that in the absence of an island, the entanglement entropy depends on the backreaction parameter in addition to linear dependence on time which influences the rising behaviour of the Page curve. Due to this, the entanglement entropy reaches the Bekenstein-Hawking entropy bound coming from the island contribution late and therefore the Page curve shifts towards a later time. In our computation of scrambling time we observe that it increases with the backreaction. 
Subsequently, we study the behaviour of mutual information in both before and after the appearance of the island scenario. We see that mutual information between the radiation subsystems vanishes at a particular observers time (critical time) before the Page time leading to a disconnected phase between the radiation subsystems. We observe that the critical time increases with backreaction parameter that signifies more correlation between the subsystems. After the Page time, the mutual information between the black hole subsystems vanishes which leads to a timescale comparable to the scrambling time.

The rest of the article is organized as follows. In \cref{sec:review} we briefly review the gravitational dual of the quark cloud model. In \cref{sec:EE} we describe the setup where the gravitational dual of the quark cloud model is coupled to thermal bath and present our computation of the entanglement entropy of radiation in the absence and presence of island. In \cref{sec-Page-curve} we study the effect of backreaction on the Page curve and scrambling time. In \cref{sec:MI} we compute and analyse the effect of backreaction on mutual information in the present setup. Finally we conclude with the summary and discussion in \cref{sec:summary}.

\section{Review of the backreacted geometry}\label{sec:review}

In this section we briefly review the gravitational dual of finite temperature strongly coupled large $N_c$ gauge theory, influenced by the presence of a uniform distribution of external heavy fundamental quarks as described in \cite{Chakrabortty:2011sp}. The correlation between the quark degrees of freedom on the boundary and the homogeneous distribution of strings in the bulk gives rise to a non-trivial deformation of the AdS-BH metric. Consider the (d+1) dimensional gravitational action corresponding to this geometry as

\begin{equation}
\label{action}
S=\frac{1}{4\pi G_{d+1}}\int dx^{d+1}\sqrt{g}\left(R-2\Lambda\right)+S_{M},
\end{equation}
where $S_{M}$ represents the matter part due to the uniform distribution of strings which is given by
\begin{equation}
\label{matteraction}
S_{M}=-\frac{1}{2}\sum_{i}\mathcal{T}_{i}\int d^2\xi \sqrt{-h}h^{\alpha\beta}\partial_{\alpha}X^{\mu}\partial_{\beta}X^{\nu}g_{\mu\nu}.
\end{equation}
Here $g_{\mu\nu}$ is the target spacetime and $h_{\alpha\beta}$ is the intrinsic metric of the string world-sheet and $\mathcal{T}_{i}$ denotes the tension of the $i$th string.
On varying the total action with respect to the space-time metric gives
\begin{equation}
R_{\mu\nu}-\frac{1}{2}R g_{\mu\nu}+\Lambda g_{\mu\nu}=8\pi G_{\mu\nu}T_{\mu\nu},
\end{equation}
where the energy-momentum tensor is given by
\begin{equation}
T^{\mu\nu}=-\sum_{i}\mathcal{T}_{i}\int{d^2\xi\frac{1}{\sqrt{|g|}}\sqrt{-h}h^{\alpha\beta}\partial_{\alpha}X^{\mu}\partial_{\beta}X^{\nu}\delta_{i}^{d-1}(x-X_i)}.
\end{equation}
One can construct the density function representing the distribution of the uniform string cloud as
\begin{equation}
\label{metric}
\nonumber b(x)=\mathcal{T}\sum_{i=1}^{N}\delta_{i}^{(d-1)}(x-X_{i}),
\end{equation}
where $N$ is the number of string and $\mathcal{T}$ is the tension of each string. Moreover, we employ the static gauge $\xi^0=t$ and $\xi^1=r$. The string density can be obtained by averaging over the $(d-1)$ spatial dimensions as

\begin{equation}
\label{metric2}
\nonumber \tilde{b}=\frac{1}{V_{d-1}}\int b(x)d^{d-1}x=\frac{\mathcal{T}N}{V_{d-1}},
\end{equation}
where $V_{d-1}$ is the volume of the $(d-1)$ dimensional space. In the limit $V_{d-1}\rightarrow \infty$, we consider very large value of $N$ to keep $N/V_{d-1}$ finite.
The non vanishing components of $T^{\mu\nu}$ are
\begin{equation}
T_{00}=-\frac{\tilde{b}}{r^3}g_{tt},~~~~~~~~~~~T_{rr}=-\frac{\tilde{b}}{r^3}g_{rr}.
\end{equation}
The ansatz for the AdS-BH metric can be expressed in the following form
\begin{equation}\label{metricd}
ds^2=-f(r)dt^2+\frac{dr^2}{f(r)}+r^2 h_{ij}dx^{i}dx^{j},
\end{equation}
where the explicit form of $f(r)$ given by
\begin{equation}
\label{V}
f(r)=K+\frac{r^2}{l^2}-\frac{2m}{r^{d-2}}-\frac{2b l^{d-3}}{(d-1)r^{d-3}},
\end{equation}
where $l$ is the AdS radius and $K$ is equal to $0,1,-1$ for the $d-1$-dimensional boundary to be flat, spherical or hyperbolic respectively. We have further redefined the string cloud density with a dimensionless parameter $b=\tilde{b}l$.

In the following sections, we will work with $K=1$ which describes the spherical symmetry in the geometry. 
 Using \cref{metricd} and \cref{V} one can compute the thermodynamical quantities by following standard holographic methods. In this work, we focus on
$d+1=4$ dimensions and we will discuss the thermodynamic quantities specifically for $d+1=4$. For a detailed stability analysis of tensor and vector perturbations, as well as more information on this model, we refer to \cite{Chakrabortty:2011sp, Chakrabortty:2016xcb}.

\section{Entanglement entropy of the Hawking radiation}
\label{sec:EE}

\subsection{Preliminaries}

\begin{figure}[ht]
	\centering
	\includegraphics[scale=1]{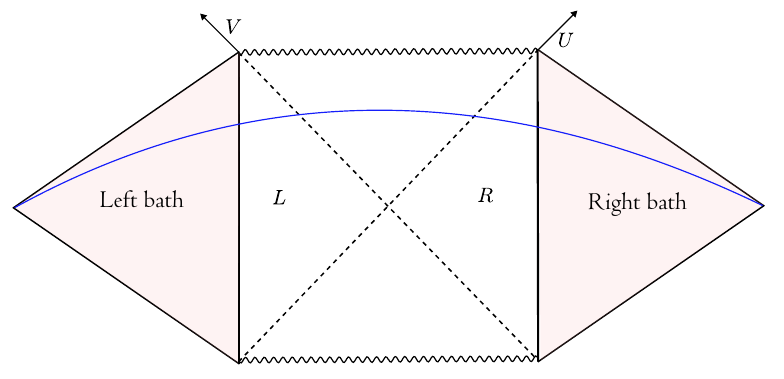}
	\caption{Schematic of the deformed eternal black hole coupled with two auxiliary flat baths. The blue curve represents the Cauchy slice that encompasses the black hole and radiation subsystems.}
	\label{set-up}
\end{figure}

As described in the previous section, the holographic bulk dual of the backreacted TFD state is described by the deformed eternal (two sided) black hole. The deformation is due to a uniform distribution of static long strings. Since the deformed black hole geometry still remains asymptotically AdS, we can attach thermal baths (non-gravitational flat spacetimes) on both sides at the asymptotic boundary of the deformed black hole with a transparent boundary condition at the interface similar to \cite{Almheiri:2019yqk}. This results in a deformed eternal black hole plus bath system in which the black hole radiates into the thermal bath and is in thermal equilibrium with it. The configuration of the present setup is depicted in \cref{set-up}. We further assume that the backreacted (deformed) black hole plus bath system is filled with conformal matter of central charge $c$ in its vacuum state. 

The metric for each side of the $(3+1)$ dimensional two-sided deformed black hole geometry is given by\footnote{For simplicity, the current computations focus on the $d=3$ case, however it can be extended to any dimension $d$.}

\begin{equation}\label{4D-metric}
ds^2=-f(r)dt^2+\frac{dr^2}{f(r)}+r^2 d\Omega^2_2,
\end{equation}
where $d\Omega_2^2=d\theta^2+\sin^2\theta d\phi^2$ and
\begin{equation}\label{f(r)}
f(r)=1+\frac{r^2}{l^2}-\frac{2m}{r}-b.
\end{equation}
The horizon radius $r_h$ are given by the real roots of $f(r_h)=0$ as

\begin{equation}
r_h=\frac{\sqrt[3]{3} (b-1) l^2+\left(\sqrt{81 l^4 m^2-3 (b-1)^3 l^6}+9 l^2 m\right)^{2/3}}{3^{2/3} \sqrt[3]{\sqrt{81 l^4 m^2-3 (b-1)^3 l^6}+9 l^2 m}},
\end{equation}
and other two roots being complex with no physical meaning. The mass of the black hole can be written in terms of $r_h$ by

\begin{equation}\label{mrh}
m=\frac{r_h \left(r_h^2+l^2-b l^2\right)}{2 l^2}.
\end{equation}
We can substitute the mass parameter $m$ from \cref{mrh} into the function given in \cref{f(r)} which reduces the number of independent parameters to two i.e $b$ and $r_h$.
The Hawking temperature and Bekenstein-Hawking entropy of the black hole are given by

\begin{align}
T_{\text{H}}&=\frac{\kappa}{2\pi}=\frac{f'(r_h)}{4\pi}=\frac{(1-b)l^2+3 r_h^2}{4\pi r_h l^2}\label{temp}\\
S_{\mathrm{BH}}&=\frac{A(r_h)}{4G_N}=\frac{\pi r_h^2}{G_N},
\end{align}
where $\kappa$ is the surface gravity of the horizon and $G_N$ is four dimensional Newton's constant. The backreaction parameter lies in the range $0\leq b\leq (1+\frac{3 r_h^2}{l^2})$ where the lower bound comes from the null energy condition of the energy momentum tensor and upper bound from the positivity of surface gravity. In the Kruskal coordinates which covers the deformed eternal black hole, the metric in \cref{4D-metric} can be written as \cite{He:2021mst}
\begin{equation}\label{Kruskal-metric}
 ds^2=-e^{2\rho}dUdV+r_{R(L)}^2d\Omega^2_2,
\end{equation}
where subscript $R$ and $L$ denotes coordinate on the right and left region of the deformed eternal black hole respectively as shown in the Penrose diagram in \cref{set-up}, and the conformal factor is given by
\begin{equation}
\begin{aligned}
\rho(r)=\frac{1}{2}\left[\log f(r_{R(L)})-2\kappa r^{*}(r_{R(L)})\right].
\end{aligned}
\end{equation}
The tortoise coordinate $r^*$ for this case can be expressed as
\begin{equation}\label{r*}
\begin{aligned}
r^*&=\int \frac{1}{f(r)}dr\\
&=\frac{l^2 r_h \sqrt{3 r_h^2-4 (b-1) l^2} \left(\log \left(1+\frac{r \left(r_h+r\right)}{r_h^2-b l^2+l^2}\right)-2 \log \left |\frac{r}{r_h}-1\right |\right)}{2 \left((b-1) l^2-3 r_h^2\right) \sqrt{3 r_h^2-4 (b-1) l^2}}\\
&\quad+\frac{l^2 \left(3 r_h^2-2 (b-1) l^2\right)}{ \left((b-1) l^2-3 r_h^2\right) \sqrt{3 r_h^2-4 (b-1) l^2}} \tan ^{-1}\left(\frac{r \sqrt{3 r_h^2-4 (b-1) l^2}}{2 (b-1) l^2-r_h \left(2 r_h+r\right)}\right).
\end{aligned}
\end{equation}
The coordinate transformations between the Kruskal coordinates and Schwarzschild coordinates in the right and left wedges of the deformed black hole are given by \cite{He:2021mst}
\begin{align}
& \mathrm{Right\, wedge}:\, &U=\kappa^{-1}\text{e}^{\kappa(t_R+r^*(r_R))},  \quad V=-\kappa^{-1}\text{e}^{-\kappa(t_R-r^*(r_R))}  \label{Kruskal-right-wedge}\\
&\mathrm{Left\, wedge}:\, &U=-\kappa^{-1}\text{e}^{-\kappa(t_L-r^*(r_L))},\quad V=\kappa^{-1}\text{e}^{\kappa(t_L+r^*(r_L))} .   \label{Kruskal-left-wedge}
\end{align}
Since the deformed black hole is glued to thermal baths on the both side where the thermal bath is a asymptotically flat region, the Kruskal coordinates can be extended to the right and left baths by using the prescription described in \cite{Almheiri:2019yqk,He:2021mst}. It involves cutting off the deformed two-sided black hole at some hypothetical distance $r_{R(L)}=\Lambda$ which lies inside the AdS boundary. The black hole and the bath metric are then smoothly joined at $r_{R(L)}=\Lambda$ by taking the normalization condition for the tortoise coordinate as $\lim\limits_{r_{R(L)}\rightarrow\infty}r_{R(L)}^*=0$.
The coordinates in the left (right) bath are given by \cite{He:2021mst}

\begin{align}
&\mathrm{Left\, bath}:\, &U=-\kappa^{-1}\text{e}^{-\kappa(t_L-r^*_L)},\quad V=\kappa^{-1}\text{e}^{\kappa(t_L+r^*_L)}  \label{Kruskal-bath-coord-left}  \\
&\mathrm{Right\, bath}:\, &U=\kappa^{-1}\text{e}^{\kappa(t_R+r^*_R)}, \quad V=-\kappa^{-1}\text{e}^{-\kappa(t_R-r^*_R)} .  \label{Kruskal-bath-coord-right}
\end{align}
Now the right and left bath corresponds to $r_{R}^*>0$ and  $r_{L}^*>0$ respectively, and $f(r)$ is given by $f(\Lambda)$ in the bath region.
With the setup described, we will now compute the fine-grained entanglement entropy of the Hawking radiation in a bath, considering the configurations with and without an island in the following subsections.

\subsection{Entanglement entropy without Island}\label{ewti}

\begin{figure}[ht]
	\centering
	\includegraphics[scale=1]{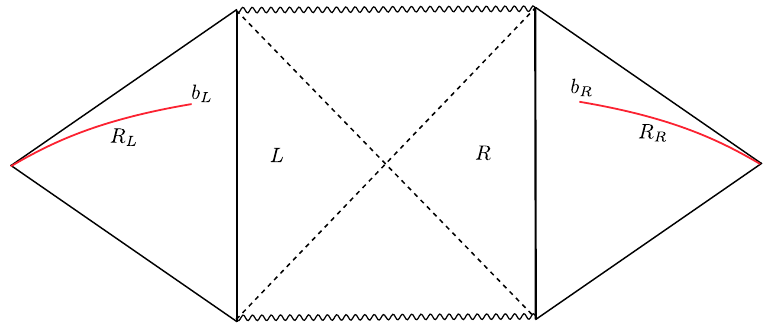}
	\caption{Schematic of the configuration having no island.}
	\label{without-island-fig}
\end{figure}

We begin by computing the radiation entanglement entropy in the absence of an island. The area term does not contribute for this case, so the entanglement entropy of the Hawking radiation is given only by the matter part of \cref{island}. The Penrose diagram of the backreacted eternal black hole coupled with thermal baths without the island configuration is shown in \cref{without-island-fig}. 

The generalized entanglement entropy can be computed by considering the von Neumann entropy of the conformal (quantum) matter on the radiation region $R_R$ and $R_L$. The boundaries of the radiation region $R_{R}$ and $R_{L}$ in the right and left baths are denoted by $b_{R}$ and $b_{L}$ respectively\footnote{Note that $b$ with a subscript e.g. $b_L$ or $b$ in subscript e.g. $r_b$ always denotes the relevant coordinate and it is not to be confused with the string density (or backreaction parameter) $b$.}. Assuming that the initial state of the total system (radiation plus black hole) is in the pure state, so the entanglement entropy of the radiation region $R\equiv R_R\cup R_L$ is equal to its complement i.e the interval $[b_L,b_R]$. The entanglement entropy of an interval having end points $x_1$ and $x_2$ is given by \cite{Almheiri:2019psf,Hashimoto:2020cas}

\begin{equation}\label{matter-entropy}
\begin{aligned}
S_{\text{matter}}(x_1,x_2)&=\frac{c}{6}\log d^2(x_1,x_2)\\
&=\frac{c}{6}\log\left\lvert\big(U(x_1)-U(x_2)\big)\big(V(x_1)-V(x_2)\big)\sqrt{\text{W}(x_1)\text{W}(x_2)}\right\lvert ,
\end{aligned}
\end{equation}
where $c$ is the central charge of the two-dimensional CFT, $d(x_1,x_2)$ is the geodesic distance between the points $x_1$ and $x_2$ in the metric of the form $ds^2=-e^{2\rho}dUdV$ and $W(x_1)$, $W(x_2)$ are the Weyl factors of the aforementioned metric. Note that \cref{matter-entropy} is valid for a two dimensional CFT. Since entanglement entropy formula for higher-dimensional space-time is not known. We can still use the formula \cref{matter-entropy} for higher-dimensional case in the s-wave approximation by assuming that location of the observer or cut-off surface $r_b$ is very far away from the black hole horizon. Therefore in the s-wave approximation we can ignore the angular part of the metric in \cref{Kruskal-metric} and can treat the radiation as CFT in flat space \cite{Hashimoto:2020cas}.

\begin{figure}[ht]
	\includegraphics[scale=0.6]{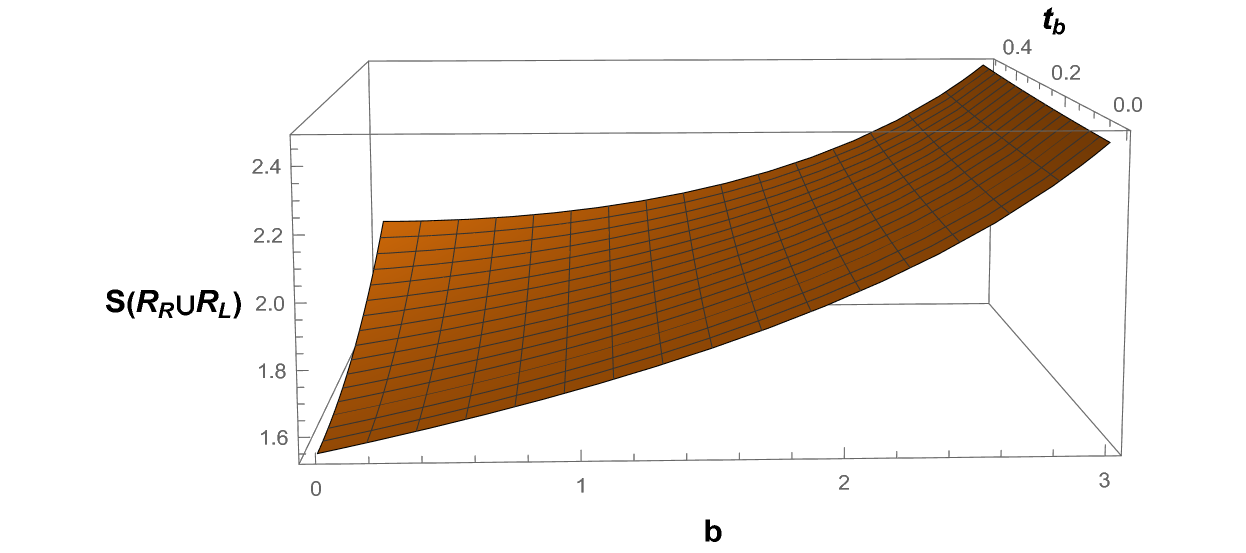}
	\caption{Plot of entanglement entropy of the Hawking radiation at early time with respect to observer time $t_b$ and string density $b$ (for $r_h=c=1, l=10$).}
	\label{etsbt}
\end{figure}

After considering the symmetric configuration $r^*_{b_L}=r^*_{b_R}=r^*_b$ and $t_{b_L}=t_{b_R}=t_b$, the entanglement entropy of the interval $[b_L,b_R]$ can be computed using the \cref{matter-entropy} as follows

\begin{equation}
S(R)=\frac{c}{6}\log\left\lvert\big(U(b_L)-U(b_{R})\big)\big(V(b_{L})-V({b_{R}})\big)\sqrt{W(b_L)W(b_R)}\right\rvert,
\end{equation}
where the Weyl factor $W$ is given by
\begin{equation}
W(b_{R(L)}) =-f(\Lambda)e^{-2\kappa r^*_b}.
\end{equation}
Now using the \cref{Kruskal-bath-coord-right} and \cref{Kruskal-bath-coord-left}, we get entanglement entropy of the Hawking radiation for the case of no island configuration as
\begin{equation}
\label{Mir}
    S(R)=S(R_R\cup R_L)=\frac{c}{6}\log{(4\kappa^{-2}f(\Lambda) \cosh^2{\kappa t_b})}.
\end{equation}
For an early time $t_b \ll \kappa^{-1}$, the entanglement entropy of the radiation $R_R\cup R_L$ shows the following behaviour
\begin{equation}
\label{ets}
    S(R_R \cup R_L)=\frac{c}{3}\log{(2\kappa^{-1}\sqrt{{f(\Lambda)})}}+ \frac{c}{6}\kappa^2 t_b^2.
\end{equation}
The above expression can be written using equations \cref{f(r)}, \cref{mrh} and \cref{temp} as
\begin{equation}
\label{etstemp}
\begin{split}
    S(R_R \cup R_L)&=\frac{c}{3}\log{\biggl[\frac{4 r_h l^2}{(1-b)l^2+3 r_h^2}\biggr]}+ \frac{c}{6}\log{\biggl[ \frac{\Lambda^2}{l^2}-\frac{r_h^3}{\Lambda l^2}+(1-\frac{r_h}{\Lambda})(1-b)\biggr]}\\&+\frac{c}{6}\biggl[\frac{(1-b)l^2+3 r_h^2}{2 r_h l^2}\biggr]^2 t_b^2.
    \end{split}
\end{equation}
In \cref{etsbt} the early time behavior of $S(R_R \cup R_L)$ i.e. \cref{etstemp} is plotted with respect to time $t_b$ and string density $b$. The entropy of the Hawking radiation $R_R \cup R_L$ increases with $b$ and $t_b$, however for large value of $b$ the increment in entropy is almost linear in time while for smaller values of $b$, the increment in entropy is quadratic in $t_b$. This is due to the first two terms of \cref{etstemp} which becomes significant for larger $b$ hence reduces the quadratic behaviour of entropy.

In the late time limit $t_b \gg \kappa^{-1}$, the entanglement entropy of the Hawking radiation given in \cref{Mir} becomes
\begin{equation}
\begin{split}
\label{lts}
    S(R_R \cup R_L)&\approx\frac{c}{3}\log\left(\kappa^{-1}\sqrt{{f(\Lambda)}}\right)+ \frac{c}{3}\kappa t_b \\&
    =\frac{c}{3}\log{\biggl[\frac{2 r_h l^2}{(1-b)l^2+3 r_h^2}\biggr]}+ \frac{c}{6}\log{\biggl[ \frac{\Lambda^2}{l^2}-\frac{r_h^3}{\Lambda l^2}+(1-\frac{r_h}{\Lambda})(1-b)\biggr]}\\&+\frac{c}{3}\biggl[\frac{(1-b)l^2+3 r_h^2}{ 2r_h l^2}\biggr] t_b .
    \end{split}
\end{equation}

For a very large time, the last term of \cref{lts} dominates therefore one can approximate the entropy as 
\begin{equation}\label{entropy-without-island}
S(R)=
 \frac{c}{6}\frac{r_h}{l^2}\left(3+\frac{(1-b) l^2}{r_h^2}\right)t_b+...,
\end{equation}
where ellipses denote the time independent terms. From the above expression it is evident that the time dependent part of $S(R)$ receives a contribution due to the backreaction $b$ which is discussed in the later section.
We observe that the entanglement entropy of the Hawking radiation rises linearly in time without any bound in the absence of an island and becomes infinite at late times which leads to the information paradox for the deformed black hole. In the following subsection we will compute the island contribution to the entanglement entropy of the Hawking radiation and show that entanglement entropy have an upper bound which dominates after the Page time.

\subsection{Entanglement entropy with Island}
\begin{figure}[ht]
	\centering
	\includegraphics[scale=1]{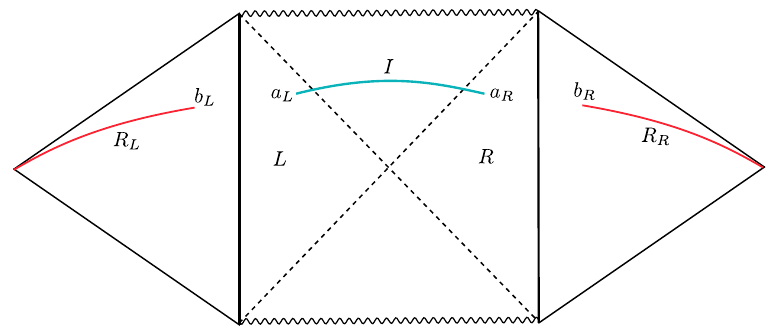}
	\caption{Schematic of the configuration in the presence of an island.}
	\label{with-island-fig}
\end{figure}

We will now show that at late time (after the Page time), an island surface emerges outside the horizon which leads to the constant value of entanglement entropy of the Hawking radiation.
Consider the configuration having an island surface as depicted in \cref{with-island-fig}. The end points of the island in the left and right wedges are labelled as $a_L$ and $a_R$ respectively. We choose the symmetric configuration such that $t_{a_L}=t_{a_R}=t_a$, $r_{a_L}=r_{a_R}=r_a$. 

Assuming that the right radiation region $R_L$ is far away from the right region $R_L$ such that the s-wave approximation can be used to determine the entanglement entropy of $R\cup I$, the von Neumann entropy of quantum (conformal) matter may be written as the sum of the $S_{\mathrm{matter}}(R\cup I)$ in the left and right regions (i.e twice the $S_{\mathrm{matter}}([a_R,b_R])$). So, the generalized entropy in the presence of an island can be obtained from \cref{island} as 

\begin{equation}
S_{\mathrm{gen}}=\frac{A(a_R)}{2G_N}+\frac{c}{3}\log\left\lvert\big(U\left({a_R}\right)-U\left({b_R}\right)\big)\big(V\left({a_R}\right)-V\left({b_R}\right)\big)\sqrt{W(a_R)W(b_R)}\right\rvert,
\end{equation}
where the Weyl factors are $W(a_R)=-f(r_a)e^{-2\kappa r^*_a}$ and $W(b_R)=-f(\Lambda)e^{-2\kappa r^*_b}$. On using \cref{Kruskal-bath-coord-right} and \cref{Kruskal-right-wedge} in above equation, we get

\begin{equation}\label{sgen-island}
S_{\mathrm{gen}}=\frac{A(r_a)}{2G_N}+\frac{c}{3}\log\bigg\lvert\frac{\sqrt{f(r_a)f(\Lambda)\text{e}^{-2\kappa(r_a^*+r^*_b)}}}{\kappa^{2}}\bigg(2\text{e}^{\kappa(r^*_a+r^*_b)}\cosh[\kappa(t_b-t_a)]-(\text{e}^{2\kappa r_a^*}+\text{e}^{2\kappa r^*_b})\bigg)\bigg\lvert.
\end{equation}
On extremizing the above \cref{sgen-island} w.r.t $t_a$ yields $t_a=t_b$ and it can now be written as

\begin{equation}\label{EE-island}
S_{\mathrm{gen}}=\frac{A(r_a)}{2G_N}+\frac{2c}{3}\log\left[\frac{\text{e}^{\kappa r^*_b}-\text{e}^{\kappa r^*_a}}{\kappa}\right]+\frac{c}{6}\log\left[ f(r_a)f(\Lambda)\text{e}^{-2\kappa\big(r^*_a+r^*_b\big)}\right].
\end{equation}
The location of QES can be determined by extremizing the above equation w.r.t $r_a$ as
\begin{equation}\label{QES-island}
\partial_{r_a}S_{\mathrm{gen}}\equiv\frac{A'(r_a)}{2G_{(D)}}-\frac{2c}{3}\frac{\kappa}{f(r_a)\bigg(\text{e}^{\kappa(r^*_b-r^*_a)}-1\bigg)}+\frac{c}{6}\frac{f'(r_a)-2\kappa}{f(r_a)}=0.
\end{equation}
The near horizon geometry for the black hole in backreacted geometry considered here is similar to other non-extremal black holes considered in \cite{He:2021mst} i.e black holes having the behaviour $f(r)\approx 2\kappa(r-r_h)$ and $r^*(r)\approx \frac{1}{2\kappa}\log\big[\frac{r}{r_h}-1\big]$ outside the horizon. The location of the island surface lies near the horizon in the limit $c\cdot G_N\ll 1$ for these general class of black holes as shown in \cite{He:2021mst}

\begin{equation}\label{ra-general}
r_a=r_h + \frac{8\kappa(c\cdot G_N)^2}{9A'(r_h)^2}\exp\Big\{-2\kappa r^*_b-2\rho(r_h)\Big\}+\mathcal{O}\left(\big(c\cdot G_N\big)^3\right).
\end{equation}
We now use \cref{f(r)} and \cref{r*} in the above \cref{ra-general} to obtain the location of the end point of the island for our case as follows

\begin{equation}\label{island-location}
\begin{aligned}
r_a&=r_h+ \frac{(c\cdot G_N)^2}{144 \pi ^2 r_h^3 \sqrt{1+\frac{2 r_h^2}{r_h^2+l^2-b l^2}}}\exp \Bigg[\frac{\left(2 (b-1) l^2-3 r_h^2\right) \tan ^{-1}\left(\frac{r_h \sqrt{3 r_h^2-4 (b-1) l^2}}{2 (b-1) l^2-3 r_h^2}\right)}{r_h \sqrt{3 r_h^2-4 (b-1) l^2}}\\
&\quad-\left(\frac{3r_h^2+(1-b) l^2}{r_h l^2}\right) r^*_b \Bigg]+\mathcal{O}\left(\big(c\cdot G_N\big)^3\right),
\end{aligned}
\end{equation}
where the area of the horizon is given by $A(r_h)=4\pi r_h^2$. On substituting the location of the island from \cref{island-location} in \cref{EE-island}, we determine the entanglement entropy of the Hawking radiation in the presence of an island, at late times as\footnote{One can also use the following formula derived in \cite{He:2021mst} for a general class of black holes having same near horizon geometry which is also applicable for our case as follows
\begin{equation}
S^{Is}(R)=\frac{A(r_h)}{2G_N}+\frac{c}{3}\log d^2(r_h,r_b)-\frac{4\kappa c^2G_N}{9A'(r_h)}\exp\big\{-2\kappa r^*_b-2\rho(r_h)\big\}+\mathcal{O}(c^3G_N^2).
\end{equation}
}

\begin{equation}\label{entropy-island}
\begin{aligned}
S(R)&=\frac{2\pi r_h^2}{G_N}+\frac{c}{3}\log d^2(r_h,r_b)-\frac{c^2 G_N }{36 \pi  r_h^2 \sqrt{1+\frac{2 r_h^2}{r_h^2+l^2-b l^2}}}\exp \Bigg[\frac{\left(2 (b-1) l^2-3 r_h^2\right) \tan ^{-1}\left(\frac{r_h \sqrt{3 r_h^2-4 (b-1) l^2}}{2 (b-1) l^2-3 r_h^2}\right)}{r_h \sqrt{3 r_h^2-4 (b-1) l^2}}\\
&\quad\quad-\left(\frac{3r_h^2+(1-b) l^2}{r_h l^2}\right) r^*_b\Bigg]+\mathcal{O}(c^3G_N^2)\\
&=2S_{\mathrm{BH}}+\mathcal{O}(c).
\end{aligned}
\end{equation}
We thus observe that the entanglement entropy of the Hawking radiation in the presence of an island saturates to a constant value i.e twice the Bekenstein-Hawking entropy of the black hole (plus sub leading order corrections in $c$) at late time.
In this way, the behavior of the entanglement entropy of the Hawking radiation after the appearance of an island follows the Page curve as discussed in the following section.

\section{Page curve and Scrambling time}\label{sec-Page-curve}

\begin{figure}[H]
\centering
  \includegraphics[width=.5\linewidth]{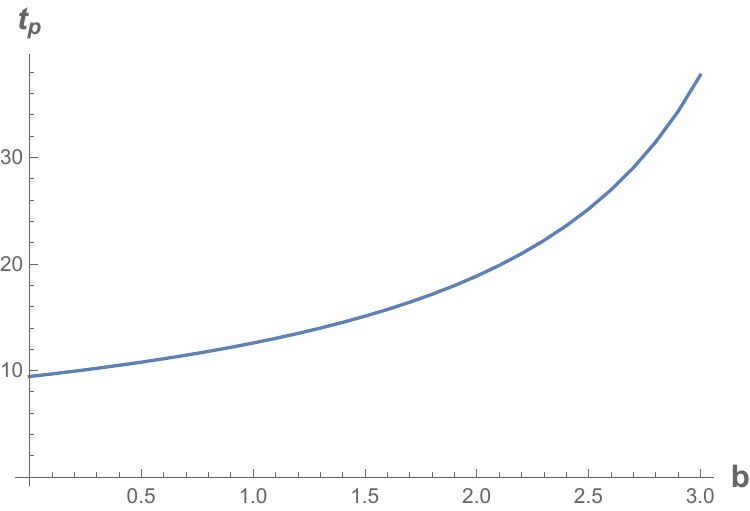}
  \caption{ Page time as a function of $b$ (for $r_h=l=c=1$).
  }
  \label{Page-time}
  \end{figure}

\begin{figure}[H]
\centering
  \includegraphics[width=.50\linewidth]{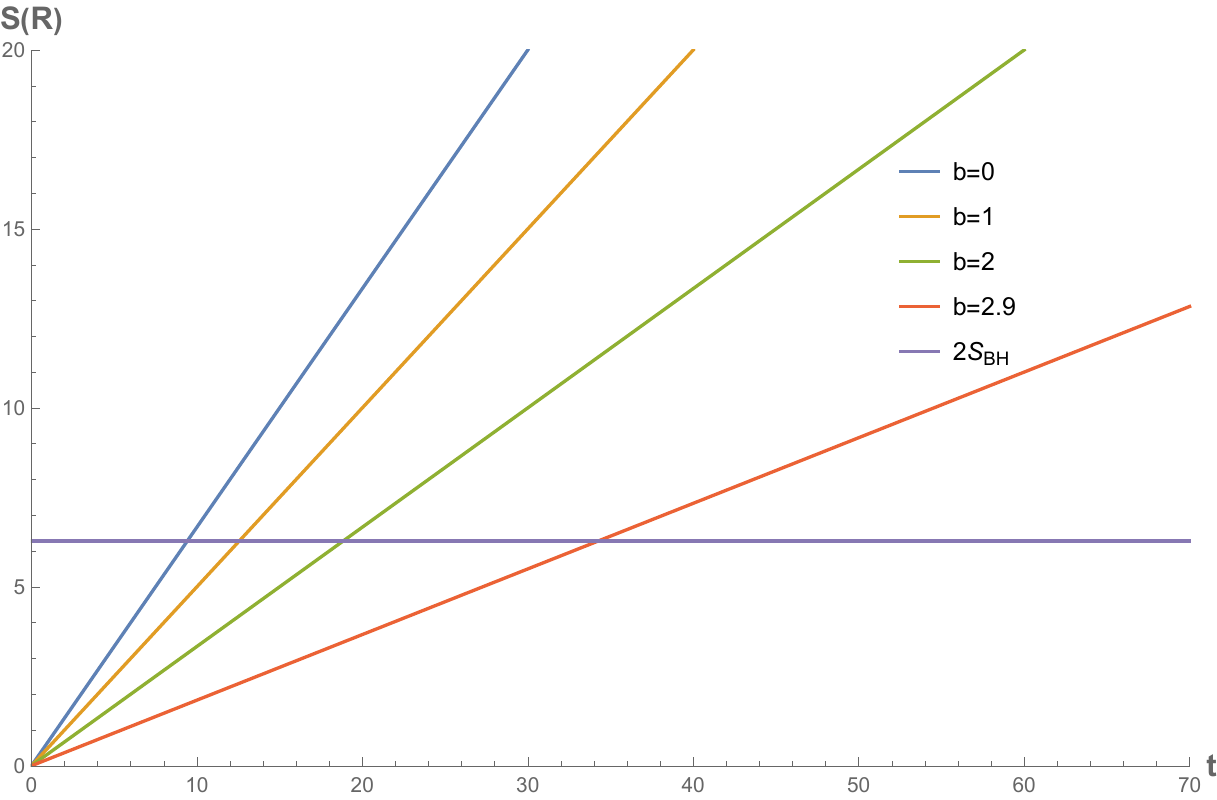}
  \caption{$S(R)$ for different values of $b$ (for $r_h=c=1, l=10$). The Page curve is given by the minimum of without island and with island contribution to the entanglement entropy of radiation.
  }
  \label{Page-curve}
  \end{figure}

Initially, the entanglement entropy of the Hawking radiation increases linearly with time due to the dominance of the configuration without the island. However due to the emergence of an island surface near the horizon at late times, the island configuration dominates and entanglement entropy becomes constant and saturates to a plateau. Thus, we get the expected unitary Page curve for a deformed eternal black hole.
The time at which the entanglement entropy reaches a constant value for a backreacted eternal black hole is called Page time. The Page time $t_P$ can be obtained by equating \cref{entropy-without-island} and \cref{entropy-island}, which may be expressed in the leading order as

\begin{equation}
t_P \sim\frac{6S_{\mathrm{BH}}}{c\kappa}=\frac{12r_h l^2}{c \left(3 r_h^2+(1-b)l^2\right)} S_{\mathrm{BH}}.
\end{equation}
It can be inferred from the above equation that Page time changes with the backreaction parameter $b$ for this model, therefore affects the emergence of island. This dependence is plotted in \cref{Page-time}.
It suggests that the non-zero $b$ makes the evaporation of the deformed black hole slow to reach the Page time compared to the AdS black hole case ($b=0$). We have plotted the Page curve for different values of $b$ in \cref{Page-curve}.

We obtain different Page curves for different values of $b$ and observe that the growth rate of entanglement entropy of the Hawking radiation in the absence of island decreases on increasing $b$. Therefore, the Page curves get shifted towards a later time indicating an increase in the Page time. It is expected because an increase in $b$ reduces the temperature of the deformed black hole which leads to the deformed black hole emitting the Hawking radiation more slowly and thus the entanglement entropy of radiation reaches the Bekenstein-Hawking entropy bound later. This causes the island which dominates after the Page time to appear late.
Note that we recover the usual Page curve for the AdS black hole in the limit $b\to 0$ which is depicted by the blue curve in \cref{Page-curve}. It serves as a consistency check for our analysis.

\subsection{Scrambling time} We now discuss the scrambling time which is defined as the shortest time interval needed to retrieve information from the Hawking radiation in the Hayden-Preskill experiment \cite{Hayden:2007cs}. This time interval represents the period during which information sent into the black hole is recoverable. 
Moreover, it is also defined as the time it takes for information to reach the boundary of the island as described by the entanglement wedge reconstruction proposal \cite{Penington:2019npb}.
To send information from the cutoff surface at $r=r_b$ into the black hole, the time required for the information to reach the boundary of the island at $r=r_a$ may be expressed as

\begin{equation}\label{scrambling-time}
\begin{aligned}
t_{\mathrm{scr}}&=r_b^*-r_a^*\\
&\simeq \frac{2 r_h l^2}{(3r_h^2+(1-b)l^2)}\log \left(\frac{\pi r_h^2}{G_N\, c}\right)+... \\
&\simeq \frac{1}{2 \pi T_{\mathrm{H}}}\log S_{\mathrm{BH}}+ \mathrm{subleading\, terms...},
\end{aligned}
\end{equation}
where \cref{island-location} and \cref{temp} are used for the location of the island and temperature of the deformed black hole respectively. We also assumed that $c\ll S_{\mathrm{BH}}$ and $r_h$, $r_b$ have the same magnitude in the above equation. The scrambling time is given by the logarithmic of Bekenstein-Hawking entropy in the leading order and resembles with the results described in \cite{Sekino:2008he}. We observe that the scrambling time of the system considered here is influenced by the deformation. As the temperature of the deformed black hole is dependent on the backreaction parameter, the scrambling time increases with increasing values of $b$. This implies that information recovery will take more time due to the deformation which conforms to our expectation that deformation delays the emergence of islands and consequently prolongs the information retrieval process.
\section{Mutual Information}\label{sec:MI}

We now move on towards the analysis of another bipartite entanglement measure known as the mutual information (MI) in our setup. The mutual information $I(A,B)$ between two subsystems $A$ and $B$ of a bipartite system $A\cup B$ is defined by $I(A,B)=S_A + S_B-S_{A\cup B}$ where $S_A$, $S_B$ and $S_{A\cup B}$ are the entanglement entropy of $A$, $B$ and $A\cup B$ respectively. It is a measure of both classical and quantum correlations between the subsystems. In the following subsections, we will compute the mutual information for both before and after Page time regimes and analyze its dependence on the backreaction (string density) parameter $b$.

\subsection{Mutual Information without Island}

\begin{figure}[H]
	\centering
	\includegraphics[scale=0.5]{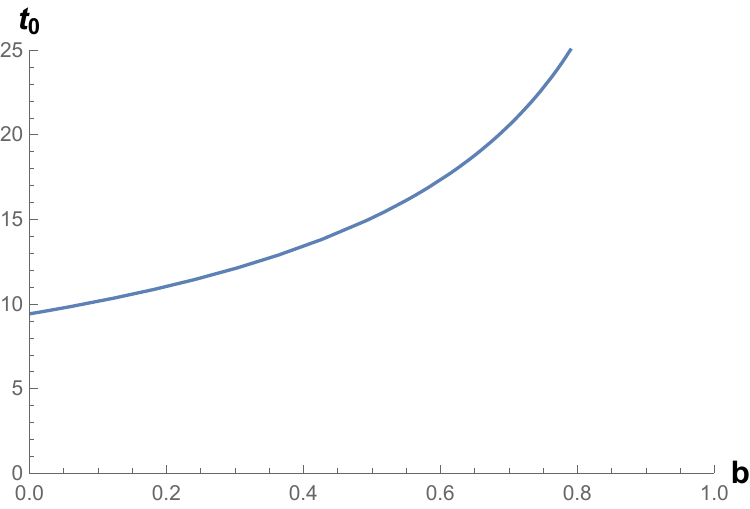}
	\caption{Plot of $t_0$ with respect to string density $b$ (for $r_h=1, l=10$).}
	\label{tob}
\end{figure}

As we have already discussed in \cref{ewti} that before the Page time, the fine grained entropy of the Hawking radiation is identified with the von Neumann entropy of matter fields in the bath region i.e. $S(R)= S_{\mathrm{matter}}(R_R\cup R_L)$. Here $R_R$ and $R_L$ are two disjoint subsystems with end points $[b_R,e_R]$ and $[b_L, e_L]$ respectively. These radiation subsystems are extended up to spatial infinity as can be seen from the \cref{without-island-fig}, with $r^*_{e_{R/L}}=r^*_e$ at $t_{e_{R/L}}=0$ where $e\to \infty$ is taken at the end of computations. The entanglement entropy of radiation subsystems $R\equiv R_R\cup R_L$ in the absence of island is given by \cref{Mir}. 
The mutual information between the radiation subsystems $R_R$ and $R_L$ is given by
\begin{equation}
    I(R_R:R_L) = S(R_R)+S(R_L)- S(R_R\cup R_L),
\end{equation}
where the expressions for $S(R_R)$ and $S(R_L)$ may be written using \cref{matter-entropy} as
\begin{equation}\label{Entropy-SR}
\begin{split}
    & S(R_R)= \frac{c}{6}\log{d^2(b_R, e_R)} 
    \\&S(R_L)= \frac{c}{6}\log{d^2(b_L, e_L)}.
    \end{split}
\end{equation}
Note that $S(R_R)= S(R_L)$ due to the artifact of $d(b_R, e_R)=d(b_L, e_L)$. Now the MI between the radiation subsystems can be obtained using \cref{Entropy-SR} and \cref{Mir} as
\begin{equation}\label{MI-before}
\begin{split}
    I(R_R:R_L) 
    &=\frac{c}{3}\log\left[2\kappa^{-2} f(\Lambda)|(\cosh{\kappa r^*_b-\cosh{\kappa t_b}})|\right]- \frac{c}{3}\log{\left[2\kappa^{-1} \sqrt{f(\Lambda)}\cosh{\kappa t_b}\right]}
    \\&= \frac{c}{3}\log\left[\kappa^{-1} \sqrt{f(\Lambda)}\biggl(\frac{\cosh{\kappa r^*_b-\cosh{\kappa t_b}}}{\cosh{\kappa t_b}}\biggr)\right].
    \end{split}
\end{equation}
For the early time $t_b \ll\kappa^{-1}$, the above expression for MI becomes
\begin{equation}
    I(R_R: R_L)= \frac{c}{3}\left[ \log{\left(\kappa^{-1}\sqrt{f(\Lambda)}\cosh{\kappa r^*_b}\right)}-\sech{\kappa r^*_b}-\frac{\kappa^2}{2}(1+\sech{\kappa r^*_b})t_b^2 \right].
\end{equation}
We note that in the early time regime, the $I(R_R:R_L)$ decreases with the time as $\sim t_b^2$. Similarly, the MI at the late time regime $t_b\gg\kappa^{-1}$ is given by
\begin{equation}
    I(R_R: R_L)= \frac{c}{3}\left[ \log{\left(\kappa^{-1}\sqrt{f(\Lambda)}\right)}-2\cosh{\kappa r^*_b}~e^{-\kappa t_b} \right],
\end{equation}
which suggests that MI increases with time as $\sim t_b$. This behavior of $I(R_R: R_L)$ at early and late times before the Page time is similar with the other models considered in \cite{RoyChowdhury:2022awr,RoyChowdhury:2023eol}. Since the MI is decreasing in earlier time and increasing at later time, there exists a critical time $t_b=t_0$ at which the MI vanishes i.e. $I(R_R:R_L)|_{t_b=t_0}=0$. This time $t_0$ can be obtained using \cref{MI-before} as

\begin{equation}
	t_0=\kappa^{-1}\cosh^{-1}\left[\frac{\kappa^{-1}\sqrt{f(\Lambda)}\cosh \kappa r^*_b}{1+\kappa^{-1}\sqrt{f(\Lambda)}}\right].
\end{equation}
The above time $t_0$ lies in the early time regime due to being $t_0\ll \kappa^{-1}$. The behavior of this critical time $t_0$ with $b$ is shown in \cref{tob}.
 We observe that the critical time increases with the increase in string density which is similar to the behavior of the Page time. 
In other words an increase in string density $b$ introduces more correlations between $R_R$ and $R_L$ hence it requires more time for the mutual information to vanish.
The behavior of mutual information before the Page time implies that the entanglement wedge of $R_R\cup R_L$ is in connected phase during the early time since $I(R_R: R_L) \neq 0$, and becomes disconnected at $t_0$ when the MI vanishes. This observation of MI is consistent with the proposal of mutual information proposed in \cite{RoyChowdhury:2022awr}.

\subsection{Mutual Information with Island}

\begin{figure}[ht]
	\centering
	\includegraphics[scale=1]{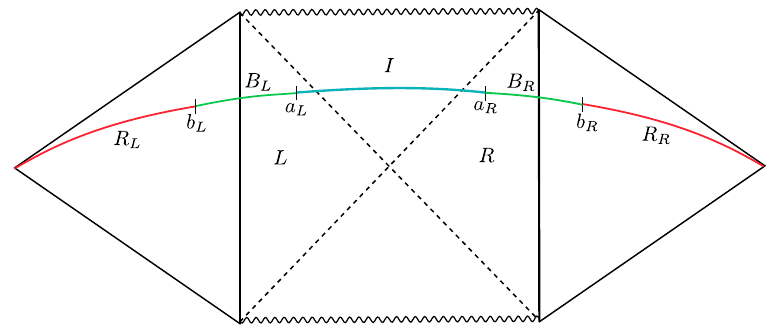}
	\caption{The black hole subsystems $B_R$ and $B_L$ are depicted by green segments.}
	\label{MI-with-island-fig}
\end{figure}

We next consider the case when the island surface emerges after the Page time and its contribution starts dominating. This configuration is depicted in \cref{MI-with-island-fig}. 
In this configuration, the entanglement entropy of radiation $S(R\equiv R_R\cup R_L)$ is given by \cref{island}. Since the whole Cauchy slice $R\cup I\cup B_R \cup B_L$ is in a pure state, we have

\begin{equation}
S_{\mathrm{matter}}(R_R\cup R_L \cup I)=S_{\mathrm{matter}}(B_R\cup B_L).
\end{equation}
The von Neumann entropy $S_{\mathrm{matter}}(B_R\cup B_L)$ of quantum (conformal) matter in the s-wave approximation may be obtained as\footnote{The von Neumann entropy of two disjoint intervals in the s-wave approximation is given by \cite{Hashimoto:2020cas} 
\begin{equation}\label{two-interval}
S_{\mathrm{matter}}(B_R\cup B_L)=\frac{c}{3}\log\Big[\frac{d(a_R,a_L)d(b_R,b_L)d(a_R,b_R)d(a_L,b_L)}{d(a_R,b_L)d(a_L,b_R)}\Big].
\end{equation}}

\begin{equation}\label{MI-BLR}
S_{\mathrm{matter}}(B_R\cup B_L) \approx S_{\mathrm{matter}}(B_R)+ S_{\mathrm{matter}}(B_L),	
\end{equation}
where $\mathcal{O}(e^{-\frac{2\pi t_{a,b}}{\beta}})$ contributions are ignored in the late time limit \cite{Hashimoto:2020cas,Matsuo:2020ypv}. The above observation is similar to the one observed for different gravitational setups in \cite{Saha:2021ohr,RoyChowdhury:2022awr,RoyChowdhury:2023eol}. So, we now use the proposal described in \cite{Saha:2021ohr} where the authors argue that the mutual information between $B_R$ and $B_L$ after the Page time vanishes i.e.

\begin{equation}\label{MI-proposal-BLR}
I(B_R:B_L)=0.
\end{equation}
It is proposed in \cite{Saha:2021ohr} that the MI between the matter fields on $B_R$ and $B_L$ vanish in the presence of an island in order to avoid the time-dependent form of $S(R)$ after the Page time. This implies that the entanglement wedge of $B_R\cup B_L$ becomes disconnected and it occurs after the Page time where the island effectively separates the entanglement wedge of $B_R\cup B_L$.
Now the proposal in \cref{MI-proposal-BLR} can be rewritten as
\begin{equation}
S_{\mathrm{matter}}(B_R\cup B_L) = S_{\mathrm{matter}}(B_R)+ S_{\mathrm{matter}}(B_L)
\end{equation}
On utilizing the \cref{matter-entropy} and \cref{two-interval} in the above equation, we obtain the following relation

\begin{equation}\label{ta-tb}
t_a-t_b=|r^*_a-r^*_b|.
\end{equation}
Now substituting the location of island from \cref{island-location} in the above equation gives

\begin{equation}\label{t-scramble-MI}
	t_a-t_b= \frac{1}{2 \pi T_{\mathrm{H}}}\log S_{\mathrm{BH}}+...=t_{\mathrm{scr}},
\end{equation}
where $t_{\mathrm{scr}}$ is the scrambling time described in \cref{scrambling-time}. 
We observe that the condition for saturation of the mutual information between black hole subsystems aligns with a time difference on the order of the scrambling time as previously noted in \cite{Saha:2021ohr,RoyChowdhury:2022awr,RoyChowdhury:2023eol}. This indicates that the emergence of island after Page time results in a disconnected entanglement wedge for $B_R\cup B_L$, described by the condition specified in \cref{t-scramble-MI}. As described earlier scrambling time is directly proportional to the backreaction $b$, we see that the saturation of the MI between black hole systems is influenced accordingly, with an increase in time difference corresponding to larger values of $b$. It is expected since deformation delays the appearance of island and thus it takes more time for island to separate the entanglement wedge of black hole subsystems.

\section{Summary and Discussion}\label{sec:summary}

To summarize, we have computed the entanglement entropy of radiation in a system where the backreacted eternal black hole is coupled to auxiliary thermal baths. In our observation we noted that the entanglement entropy of the Hawking radiation receives a significant contribution due the presence of backreaction parameter $b$. This backreaction parameter corresponds to the deformation of the black hole in the bulk gravitational theory. The island formula was utilized to obtain the entanglement entropy of the Hawking radiation in baths coming through the AdS boundaries on which transparent boundary conditions are imposed. First we computed the entanglement entropy of radiation in the absence of an island using the s-wave approximation in the early and late time limit. We found that it shows a quadraticaly increasing behaviour with time that deviates slightly from quadratic nature as the value of backreaction parameter increased. In the late time we observe a linear increment in entropy with respect to time as well as with backreaction parameter. Then we obtained the entanglement entropy of radiation in the presence of an island. It is observed that the entanglement entropy for the island configuration gives twice the Bekenstein-Hawking entropy of the deformed black hole at the leading order and is independent of time. 
Hence, we observed that the entanglement entropy of radiation increases during the initial period of the black hole evaporation and at the later stage, an island appears slightly outside the horizon which leads to a constant value of entanglement entropy and we get the Page curve.

Subsequently we studied the impact of the backreaction parameter on the entanglement entropy of radiation and its effect on the Page curve. It was observed that the Page curve depends on the backreaction parameter and gets shifted towards a later time as the value of the backreaction parameter is increased. This is because backreaction decreases the temperature of the deformed black hole causing it to evaporate slowly. As a result, the growth rate of the entanglement entropy slows down initially and it reaches the saturation value at a later time. Consequently the emergence of the island is delayed and the Page time gets shifted to a later time. In the absence of backreaction our result recover the Page curve of the eternal AdS black hole. We calculated the scrambling time and observed that it is influenced by the backreaction. This backreaction or deformation increases the scrambling time which leads to longer time duration required for the information retrieval. Furthermore, we computed the mutual information for backreacted system. Before the Page time, we noticed that the MI between the radiation subsystems is constant at the initial time and decreases as time increases. Eventually MI becomes zero at a critical value of time. This critical value of time is affected by the backreaction $b$ present in the system. We observe that the critical time and the Page time shows a similar dependency on parameter $b$. Post Page time, the emergence of an island gives way to the disconnected entanglement wedge corresponding to black hole subsystems which in turn implies vanishing of the MI between them. It was observed that the timescale associated with this MI saturation aligns with the order of scrambling time.

The backreaction parameter $b$ represents the external quarks in the quark cloud model. These external degrees of freedom introduced into the system affects the entanglement entropy of the Hawking radiation and consequently we see an increase in the Page time. Furthermore, as we increase $b$, it takes a longer time to scramble the information. This increment in the scrambling time is in expectation with the results as observed in \cite{Chakrabortty:2022kvq}. The scrambling time is also related with the saturation of the mutual information after the island formation so an increase in $b$ delays the time required for the mutual information to go to zero. Our analysis captures important and novel insights about the Page curve, scrambling time and mutual information via the island prescription as we are able to extract an exciting interplay between (i) quark density and Page time (ii) quark density and scrambling time (iii) quark density and mutual information
which has not been reported in the literature so far. These results can play an important role in understanding the relationship between strongly coupled gauge theories and black holes as it accentuates the quantum information theoretic aspects of this relationship.

In this paper, we have studied the Page curve and MI of the deformed eternal black hole, it would be interesting to investigate the behaviour of the Page curve for an evaporating black hole with deformation.  It would also be extremely interesting to obtain the entanglement entropy of radiation using the double holography picture, where the computation reduces to calculating the RT surface in higher dimensions.  Another interesting research direction could be to explore other quantum information theoretic measures such as entanglement negativity and reflected entropy, in the present gravitational setup using the island prescription to gain a better understanding of the black hole information paradox. We leave these open issues for future work.

\section{Acknowledgments}

We would like to thank the organizers and participants of \textit{QBIS 2023} at APCTP, Korea where the idea for this project was conceived.
S.P. would like to thank Shankhadeep Chakrabortty for his valuable comments and discussion. 
The work of P. J. has been supported by an appointment to the JRG Program at the APCTP through the Science and Technology Promotion Fund and Lottery Fund of the Korean Government. P. J. is also supported by the Korean Local Governments – Gyeongsangbuk-do Province and Pohang City – and by the National Research Foundation of Korea (NRF) funded by the Korean government (MSIT)(grant number 2021R1A2C1010834). S.P. acknowledges the support of Senior Research Fellowship from the Ministry of Human Resource and Development, Government of India. H.P. acknowledges the support of this work by the NCTS, Taiwan.\\

\bibliographystyle{JHEP}
\bibliography{IsQCM-ref}

\providecommand{\href}[2]{#2}\begingroup\raggedright\begin{thebibliography}{100}

\bibitem{Hawking:1975vcx}
S.W.~Hawking, \emph{{Particle Creation by Black Holes}}, \href{https://doi.org/10.1007/BF02345020}{\emph{Commun. Math. Phys.} {\bfseries 43} (1975) 199}.

\bibitem{Hawking:1976ra}
S.W.~Hawking, \emph{{Breakdown of Predictability in Gravitational Collapse}}, \href{https://doi.org/10.1103/PhysRevD.14.2460}{\emph{Phys. Rev. D} {\bfseries 14} (1976) 2460}.

\bibitem{Page:1993wv}
D.N.~Page, \emph{{Information in black hole radiation}}, \href{https://doi.org/10.1103/PhysRevLett.71.3743}{\emph{Phys. Rev. Lett.} {\bfseries 71} (1993) 3743} [\href{https://arxiv.org/abs/hep-th/9306083}{{\ttfamily hep-th/9306083}}].

\bibitem{Page:2013dx}
D.N.~Page, \emph{{Time Dependence of Hawking Radiation Entropy}}, \href{https://doi.org/10.1088/1475-7516/2013/09/028}{\emph{JCAP} {\bfseries 09} (2013) 028} [\href{https://arxiv.org/abs/1301.4995}{{\ttfamily 1301.4995}}].

\bibitem{Penington:2019npb}
G.~Penington, \emph{{Entanglement Wedge Reconstruction and the Information Paradox}}, \href{https://doi.org/10.1007/JHEP09(2020)002}{\emph{JHEP} {\bfseries 09} (2020) 002} [\href{https://arxiv.org/abs/1905.08255}{{\ttfamily 1905.08255}}].

\bibitem{Almheiri:2019psf}
A.~Almheiri, N.~Engelhardt, D.~Marolf and H.~Maxfield, \emph{{The entropy of bulk quantum fields and the entanglement wedge of an evaporating black hole}}, \href{https://doi.org/10.1007/JHEP12(2019)063}{\emph{JHEP} {\bfseries 12} (2019) 063} [\href{https://arxiv.org/abs/1905.08762}{{\ttfamily 1905.08762}}].

\bibitem{Almheiri:2019hni}
A.~Almheiri, R.~Mahajan, J.~Maldacena and Y.~Zhao, \emph{{The Page curve of Hawking radiation from semiclassical geometry}}, \href{https://doi.org/10.1007/JHEP03(2020)149}{\emph{JHEP} {\bfseries 03} (2020) 149} [\href{https://arxiv.org/abs/1908.10996}{{\ttfamily 1908.10996}}].

\bibitem{Almheiri:2019yqk}
A.~Almheiri, R.~Mahajan and J.~Maldacena, \emph{{Islands outside the horizon}},  \href{https://arxiv.org/abs/1910.11077}{{\ttfamily 1910.11077}}.

\bibitem{Almheiri:2019psy}
A.~Almheiri, R.~Mahajan and J.E.~Santos, \emph{{Entanglement islands in higher dimensions}}, \href{https://doi.org/10.21468/SciPostPhys.9.1.001}{\emph{SciPost Phys.} {\bfseries 9} (2020) 001} [\href{https://arxiv.org/abs/1911.09666}{{\ttfamily 1911.09666}}].

\bibitem{Almheiri:2020cfm}
A.~Almheiri, T.~Hartman, J.~Maldacena, E.~Shaghoulian and A.~Tajdini, \emph{{The entropy of Hawking radiation}}, \href{https://doi.org/10.1103/RevModPhys.93.035002}{\emph{Rev. Mod. Phys.} {\bfseries 93} (2021) 035002} [\href{https://arxiv.org/abs/2006.06872}{{\ttfamily 2006.06872}}].

\bibitem{Engelhardt:2014gca}
N.~Engelhardt and A.C.~Wall, \emph{{Quantum Extremal Surfaces: Holographic Entanglement Entropy beyond the Classical Regime}}, \href{https://doi.org/10.1007/JHEP01(2015)073}{\emph{JHEP} {\bfseries 01} (2015) 073} [\href{https://arxiv.org/abs/1408.3203}{{\ttfamily 1408.3203}}].

\bibitem{Ryu:2006bv}
S.~Ryu and T.~Takayanagi, \emph{{Holographic derivation of entanglement entropy from AdS/CFT}}, \href{https://doi.org/10.1103/PhysRevLett.96.181602}{\emph{Phys. Rev. Lett.} {\bfseries 96} (2006) 181602} [\href{https://arxiv.org/abs/hep-th/0603001}{{\ttfamily hep-th/0603001}}].

\bibitem{Page:1993df}
D.N.~Page, \emph{{Average entropy of a subsystem}}, \href{https://doi.org/10.1103/PhysRevLett.71.1291}{\emph{Phys. Rev. Lett.} {\bfseries 71} (1993) 1291} [\href{https://arxiv.org/abs/gr-qc/9305007}{{\ttfamily gr-qc/9305007}}].

\bibitem{Penington:2019kki}
G.~Penington, S.H.~Shenker, D.~Stanford and Z.~Yang, \emph{{Replica wormholes and the black hole interior}}, \href{https://doi.org/10.1007/JHEP03(2022)205}{\emph{JHEP} {\bfseries 03} (2022) 205} [\href{https://arxiv.org/abs/1911.11977}{{\ttfamily 1911.11977}}].

\bibitem{Almheiri:2019qdq}
A.~Almheiri, T.~Hartman, J.~Maldacena, E.~Shaghoulian and A.~Tajdini, \emph{{Replica Wormholes and the Entropy of Hawking Radiation}}, \href{https://doi.org/10.1007/JHEP05(2020)013}{\emph{JHEP} {\bfseries 05} (2020) 013} [\href{https://arxiv.org/abs/1911.12333}{{\ttfamily 1911.12333}}].

\bibitem{Hashimoto:2020cas}
K.~Hashimoto, N.~Iizuka and Y.~Matsuo, \emph{{Islands in Schwarzschild black holes}}, \href{https://doi.org/10.1007/JHEP06(2020)085}{\emph{JHEP} {\bfseries 06} (2020) 085} [\href{https://arxiv.org/abs/2004.05863}{{\ttfamily 2004.05863}}].

\bibitem{Wang:2021woy}
X.~Wang, R.~Li and J.~Wang, \emph{{Islands and Page curves of Reissner-Nordstr\"om black holes}}, \href{https://doi.org/10.1007/JHEP04(2021)103}{\emph{JHEP} {\bfseries 04} (2021) 103} [\href{https://arxiv.org/abs/2101.06867}{{\ttfamily 2101.06867}}].

\bibitem{Yu:2021cgi}
M.-H.~Yu and X.-H.~Ge, \emph{{Islands and Page curves in charged dilaton black holes}}, \href{https://doi.org/10.1140/epjc/s10052-021-09932-w}{\emph{Eur. Phys. J. C} {\bfseries 82} (2022) 14} [\href{https://arxiv.org/abs/2107.03031}{{\ttfamily 2107.03031}}].

\bibitem{Ahn:2021chg}
B.~Ahn, S.-E.~Bak, H.-S.~Jeong, K.-Y.~Kim and Y.-W.~Sun, \emph{{Islands in charged linear dilaton black holes}}, \href{https://doi.org/10.1103/PhysRevD.105.046012}{\emph{Phys. Rev. D} {\bfseries 105} (2022) 046012} [\href{https://arxiv.org/abs/2107.07444}{{\ttfamily 2107.07444}}].

\bibitem{Karananas:2020fwx}
G.K.~Karananas, A.~Kehagias and J.~Taskas, \emph{{Islands in linear dilaton black holes}}, \href{https://doi.org/10.1007/JHEP03(2021)253}{\emph{JHEP} {\bfseries 03} (2021) 253} [\href{https://arxiv.org/abs/2101.00024}{{\ttfamily 2101.00024}}].

\bibitem{Lu:2021gmv}
Y.~Lu and J.~Lin, \emph{{Islands in Kaluza\textendash{}Klein black holes}}, \href{https://doi.org/10.1140/epjc/s10052-022-10074-w}{\emph{Eur. Phys. J. C} {\bfseries 82} (2022) 132} [\href{https://arxiv.org/abs/2106.07845}{{\ttfamily 2106.07845}}].

\bibitem{Arefeva:2021kfx}
I.~Aref'eva and I.~Volovich, \emph{{A note on islands in Schwarzschild black holes}}, \href{https://doi.org/10.4213/tmf10386}{\emph{Teor. Mat. Fiz.} {\bfseries 214} (2023) 500} [\href{https://arxiv.org/abs/2110.04233}{{\ttfamily 2110.04233}}].

\bibitem{He:2021mst}
S.~He, Y.~Sun, L.~Zhao and Y.-X.~Zhang, \emph{{The universality of islands outside the horizon}}, \href{https://doi.org/10.1007/JHEP05(2022)047}{\emph{JHEP} {\bfseries 05} (2022) 047} [\href{https://arxiv.org/abs/2110.07598}{{\ttfamily 2110.07598}}].

\bibitem{Chandrasekaran:2020qtn}
V.~Chandrasekaran, M.~Miyaji and P.~Rath, \emph{{Including contributions from entanglement islands to the reflected entropy}}, \href{https://doi.org/10.1103/PhysRevD.102.086009}{\emph{Phys. Rev. D} {\bfseries 102} (2020) 086009} [\href{https://arxiv.org/abs/2006.10754}{{\ttfamily 2006.10754}}].

\bibitem{KumarBasak:2020ams}
J.~Kumar~Basak, D.~Basu, V.~Malvimat, H.~Parihar and G.~Sengupta, \emph{{Islands for entanglement negativity}}, \href{https://doi.org/10.21468/SciPostPhys.12.1.003}{\emph{SciPost Phys.} {\bfseries 12} (2022) 003} [\href{https://arxiv.org/abs/2012.03983}{{\ttfamily 2012.03983}}].

\bibitem{KumarBasak:2021rrx}
J.~Kumar~Basak, D.~Basu, V.~Malvimat, H.~Parihar and G.~Sengupta, \emph{{Page curve for entanglement negativity through geometric evaporation}}, \href{https://doi.org/10.21468/SciPostPhys.12.1.004}{\emph{SciPost Phys.} {\bfseries 12} (2022) 004} [\href{https://arxiv.org/abs/2106.12593}{{\ttfamily 2106.12593}}].

\bibitem{Bhattacharya:2021jrn}
A.~Bhattacharya, A.~Bhattacharyya, P.~Nandy and A.K.~Patra, \emph{{Islands and complexity of eternal black hole and radiation subsystems for a doubly holographic model}}, \href{https://doi.org/10.1007/JHEP05(2021)135}{\emph{JHEP} {\bfseries 05} (2021) 135} [\href{https://arxiv.org/abs/2103.15852}{{\ttfamily 2103.15852}}].

\bibitem{Li:2021dmf}
T.~Li, M.-K.~Yuan and Y.~Zhou, \emph{{Defect extremal surface for reflected entropy}}, \href{https://doi.org/10.1007/JHEP01(2022)018}{\emph{JHEP} {\bfseries 01} (2022) 018} [\href{https://arxiv.org/abs/2108.08544}{{\ttfamily 2108.08544}}].

\bibitem{Bhattacharya:2021dnd}
A.~Bhattacharya, A.~Bhattacharyya, P.~Nandy and A.K.~Patra, \emph{{Partial islands and subregion complexity in geometric secret-sharing model}}, \href{https://doi.org/10.1007/JHEP12(2021)091}{\emph{JHEP} {\bfseries 12} (2021) 091} [\href{https://arxiv.org/abs/2109.07842}{{\ttfamily 2109.07842}}].

\bibitem{Ling:2021vxe}
Y.~Ling, P.~Liu, Y.~Liu, C.~Niu, Z.-Y.~Xian and C.-Y.~Zhang, \emph{{Reflected entropy in double holography}}, \href{https://doi.org/10.1007/JHEP02(2022)037}{\emph{JHEP} {\bfseries 02} (2022) 037} [\href{https://arxiv.org/abs/2109.09243}{{\ttfamily 2109.09243}}].

\bibitem{Saha:2021ohr}
A.~Saha, S.~Gangopadhyay and J.P.~Saha, \emph{{Mutual information, islands in black holes and the Page curve}}, \href{https://doi.org/10.1140/epjc/s10052-022-10426-6}{\emph{Eur. Phys. J. C} {\bfseries 82} (2022) 476} [\href{https://arxiv.org/abs/2109.02996}{{\ttfamily 2109.02996}}].

\bibitem{Akers:2022max}
C.~Akers, T.~Faulkner, S.~Lin and P.~Rath, \emph{{The Page curve for reflected entropy}}, \href{https://doi.org/10.1007/JHEP06(2022)089}{\emph{JHEP} {\bfseries 06} (2022) 089} [\href{https://arxiv.org/abs/2201.11730}{{\ttfamily 2201.11730}}].

\bibitem{BasakKumar:2022stg}
J.~Basak~Kumar, D.~Basu, V.~Malvimat, H.~Parihar and G.~Sengupta, \emph{{Reflected entropy and entanglement negativity for holographic moving mirrors}}, \href{https://doi.org/10.1007/JHEP09(2022)089}{\emph{JHEP} {\bfseries 09} (2022) 089} [\href{https://arxiv.org/abs/2204.06015}{{\ttfamily 2204.06015}}].

\bibitem{Lin:2022qfn}
J.~Lin and Y.~Lu, \emph{{Effective reflected entropy and entanglement negativity for general 2D eternal black holes}},  \href{https://arxiv.org/abs/2204.08290}{{\ttfamily 2204.08290}}.

\bibitem{Basu:2022reu}
D.~Basu, H.~Parihar, V.~Raj and G.~Sengupta, \emph{{Defect extremal surfaces for entanglement negativity}},  \href{https://arxiv.org/abs/2205.07905}{{\ttfamily 2205.07905}}.

\bibitem{Shao:2022gpg}
Y.~Shao, M.-K.~Yuan and Y.~Zhou, \emph{{Entanglement Negativity and Defect Extremal Surface}},  \href{https://arxiv.org/abs/2206.05951}{{\ttfamily 2206.05951}}.

\bibitem{Afrasiar:2022ebi}
M.~Afrasiar, J.~Kumar~Basak, A.~Chandra and G.~Sengupta, \emph{{Islands for entanglement negativity in communicating black holes}}, \href{https://doi.org/10.1103/PhysRevD.108.066013}{\emph{Phys. Rev. D} {\bfseries 108} (2023) 066013} [\href{https://arxiv.org/abs/2205.07903}{{\ttfamily 2205.07903}}].

\bibitem{RoyChowdhury:2022awr}
A.~Roy~Chowdhury, A.~Saha and S.~Gangopadhyay, \emph{{Role of mutual information in the Page curve}}, \href{https://doi.org/10.1103/PhysRevD.106.086019}{\emph{Phys. Rev. D} {\bfseries 106} (2022) 086019} [\href{https://arxiv.org/abs/2207.13029}{{\ttfamily 2207.13029}}].

\bibitem{Afrasiar:2022fid}
M.~Afrasiar, J.K.~Basak, A.~Chandra and G.~Sengupta, \emph{{Reflected entropy for communicating black holes. Part I. Karch-Randall braneworlds}}, \href{https://doi.org/10.1007/JHEP02(2023)203}{\emph{JHEP} {\bfseries 02} (2023) 203} [\href{https://arxiv.org/abs/2211.13246}{{\ttfamily 2211.13246}}].

\bibitem{Afrasiar:2023jrj}
M.~Afrasiar, J.K.~Basak, A.~Chandra and G.~Sengupta, \emph{{Reflected Entropy for Communicating Black Holes II: Planck Braneworlds}},  \href{https://arxiv.org/abs/2302.12810}{{\ttfamily 2302.12810}}.

\bibitem{RoyChowdhury:2023eol}
A.~Roy~Chowdhury, A.~Saha and S.~Gangopadhyay, \emph{{Mutual information of subsystems and the Page curve for the Schwarzschild\textendash{}de Sitter black hole}}, \href{https://doi.org/10.1103/PhysRevD.108.026003}{\emph{Phys. Rev. D} {\bfseries 108} (2023) 026003} [\href{https://arxiv.org/abs/2303.14062}{{\ttfamily 2303.14062}}].

\bibitem{Kumari:2023ops}
A.~Kumari, V.~Raj and G.~Sengupta, \emph{{Odd entanglement entropy in boundary conformal field theories and holographic moving mirrors}},  \href{https://arxiv.org/abs/2310.11242}{{\ttfamily 2310.11242}}.

\bibitem{Goto:2020wnk}
K.~Goto, T.~Hartman and A.~Tajdini, \emph{{Replica wormholes for an evaporating 2D black hole}}, \href{https://doi.org/10.1007/JHEP04(2021)289}{\emph{JHEP} {\bfseries 04} (2021) 289} [\href{https://arxiv.org/abs/2011.09043}{{\ttfamily 2011.09043}}].

\bibitem{Colin-Ellerin:2020mva}
S.~Colin-Ellerin, X.~Dong, D.~Marolf, M.~Rangamani and Z.~Wang, \emph{{Real-time gravitational replicas: Formalism and a variational principle}}, \href{https://doi.org/10.1007/JHEP05(2021)117}{\emph{JHEP} {\bfseries 05} (2021) 117} [\href{https://arxiv.org/abs/2012.00828}{{\ttfamily 2012.00828}}].

\bibitem{Kawabata:2021vyo}
K.~Kawabata, T.~Nishioka, Y.~Okuyama and K.~Watanabe, \emph{{Replica wormholes and capacity of entanglement}}, \href{https://doi.org/10.1007/JHEP10(2021)227}{\emph{JHEP} {\bfseries 10} (2021) 227} [\href{https://arxiv.org/abs/2105.08396}{{\ttfamily 2105.08396}}].

\bibitem{Geng:2020qvw}
H.~Geng and A.~Karch, \emph{{Massive islands}}, \href{https://doi.org/10.1007/JHEP09(2020)121}{\emph{JHEP} {\bfseries 09} (2020) 121} [\href{https://arxiv.org/abs/2006.02438}{{\ttfamily 2006.02438}}].

\bibitem{Geng:2020fxl}
H.~Geng, A.~Karch, C.~Perez-Pardavila, S.~Raju, L.~Randall, M.~Riojas et~al., \emph{{Information Transfer with a Gravitating Bath}}, \href{https://doi.org/10.21468/SciPostPhys.10.5.103}{\emph{SciPost Phys.} {\bfseries 10} (2021) 103} [\href{https://arxiv.org/abs/2012.04671}{{\ttfamily 2012.04671}}].

\bibitem{Deng:2020ent}
F.~Deng, J.~Chu and Y.~Zhou, \emph{{Defect extremal surface as the holographic counterpart of Island formula}}, \href{https://doi.org/10.1007/JHEP03(2021)008}{\emph{JHEP} {\bfseries 03} (2021) 008} [\href{https://arxiv.org/abs/2012.07612}{{\ttfamily 2012.07612}}].

\bibitem{Krishnan:2020oun}
C.~Krishnan, V.~Patil and J.~Pereira, \emph{{Page Curve and the Information Paradox in Flat Space}},  \href{https://arxiv.org/abs/2005.02993}{{\ttfamily 2005.02993}}.

\bibitem{Krishnan:2020fer}
C.~Krishnan, \emph{{Critical Islands}}, \href{https://doi.org/10.1007/JHEP01(2021)179}{\emph{JHEP} {\bfseries 01} (2021) 179} [\href{https://arxiv.org/abs/2007.06551}{{\ttfamily 2007.06551}}].

\bibitem{Hollowood:2020cou}
T.J.~Hollowood and S.P.~Kumar, \emph{{Islands and Page Curves for Evaporating Black Holes in JT Gravity}}, \href{https://doi.org/10.1007/JHEP08(2020)094}{\emph{JHEP} {\bfseries 08} (2020) 094} [\href{https://arxiv.org/abs/2004.14944}{{\ttfamily 2004.14944}}].

\bibitem{Chen:2020uac}
H.Z.~Chen, R.C.~Myers, D.~Neuenfeld, I.A.~Reyes and J.~Sandor, \emph{{Quantum Extremal Islands Made Easy, Part I: Entanglement on the Brane}}, \href{https://doi.org/10.1007/JHEP10(2020)166}{\emph{JHEP} {\bfseries 10} (2020) 166} [\href{https://arxiv.org/abs/2006.04851}{{\ttfamily 2006.04851}}].

\bibitem{Chen:2020hmv}
H.Z.~Chen, R.C.~Myers, D.~Neuenfeld, I.A.~Reyes and J.~Sandor, \emph{{Quantum Extremal Islands Made Easy, Part II: Black Holes on the Brane}}, \href{https://doi.org/10.1007/JHEP12(2020)025}{\emph{JHEP} {\bfseries 12} (2020) 025} [\href{https://arxiv.org/abs/2010.00018}{{\ttfamily 2010.00018}}].

\bibitem{Hernandez:2020nem}
J.~Hernandez, R.C.~Myers and S.-M.~Ruan, \emph{{Quantum extremal islands made easy. Part III. Complexity on the brane}}, \href{https://doi.org/10.1007/JHEP02(2021)173}{\emph{JHEP} {\bfseries 02} (2021) 173} [\href{https://arxiv.org/abs/2010.16398}{{\ttfamily 2010.16398}}].

\bibitem{Akal:2020twv}
I.~Akal, Y.~Kusuki, N.~Shiba, T.~Takayanagi and Z.~Wei, \emph{{Entanglement Entropy in a Holographic Moving Mirror and the Page Curve}}, \href{https://doi.org/10.1103/PhysRevLett.126.061604}{\emph{Phys. Rev. Lett.} {\bfseries 126} (2021) 061604} [\href{https://arxiv.org/abs/2011.12005}{{\ttfamily 2011.12005}}].

\bibitem{Uhlemann:2021nhu}
C.F.~Uhlemann, \emph{{Islands and Page curves in 4d from Type IIB}}, \href{https://doi.org/10.1007/JHEP08(2021)104}{\emph{JHEP} {\bfseries 08} (2021) 104} [\href{https://arxiv.org/abs/2105.00008}{{\ttfamily 2105.00008}}].

\bibitem{Yu:2021rfg}
M.-H.~Yu, C.-Y.~Lu, X.-H.~Ge and S.-J.~Sin, \emph{{Island, Page curve, and superradiance of rotating BTZ black holes}}, \href{https://doi.org/10.1103/PhysRevD.105.066009}{\emph{Phys. Rev. D} {\bfseries 105} (2022) 066009} [\href{https://arxiv.org/abs/2112.14361}{{\ttfamily 2112.14361}}].

\bibitem{Krishnan:2021faa}
C.~Krishnan and V.~Mohan, \emph{{Hints of gravitational ergodicity: Berry\textquoteright{}s ensemble and the universality of the semi-classical Page curve}}, \href{https://doi.org/10.1007/JHEP05(2021)126}{\emph{JHEP} {\bfseries 05} (2021) 126} [\href{https://arxiv.org/abs/2102.07703}{{\ttfamily 2102.07703}}].

\bibitem{Chu:2021gdb}
J.~Chu, F.~Deng and Y.~Zhou, \emph{{Page curve from defect extremal surface and island in higher dimensions}}, \href{https://doi.org/10.1007/JHEP10(2021)149}{\emph{JHEP} {\bfseries 10} (2021) 149} [\href{https://arxiv.org/abs/2105.09106}{{\ttfamily 2105.09106}}].

\bibitem{Ghosh:2021axl}
K.~Ghosh and C.~Krishnan, \emph{{Dirichlet baths and the not-so-fine-grained Page curve}}, \href{https://doi.org/10.1007/JHEP08(2021)119}{\emph{JHEP} {\bfseries 08} (2021) 119} [\href{https://arxiv.org/abs/2103.17253}{{\ttfamily 2103.17253}}].

\bibitem{Krishnan:2021ffb}
C.~Krishnan and V.~Mohan, \emph{{Interpreting the Bulk Page Curve: A Vestige of Locality on Holographic Screens}},  \href{https://arxiv.org/abs/2112.13783}{{\ttfamily 2112.13783}}.

\bibitem{Hollowood:2021nlo}
T.J.~Hollowood, S.P.~Kumar, A.~Legramandi and N.~Talwar, \emph{{Islands in the stream of Hawking radiation}}, \href{https://doi.org/10.1007/JHEP11(2021)067}{\emph{JHEP} {\bfseries 11} (2021) 067} [\href{https://arxiv.org/abs/2104.00052}{{\ttfamily 2104.00052}}].

\bibitem{Hollowood:2021wkw}
T.J.~Hollowood, S.P.~Kumar, A.~Legramandi and N.~Talwar, \emph{{Ephemeral islands, plunging quantum extremal surfaces and BCFT channels}}, \href{https://doi.org/10.1007/JHEP01(2022)078}{\emph{JHEP} {\bfseries 01} (2022) 078} [\href{https://arxiv.org/abs/2109.01895}{{\ttfamily 2109.01895}}].

\bibitem{Akal:2021dqt}
I.~Akal, T.~Kawamoto, S.-M.~Ruan, T.~Takayanagi and Z.~Wei, \emph{{Page curve under final state projection}}, \href{https://doi.org/10.1103/PhysRevD.105.126026}{\emph{Phys. Rev. D} {\bfseries 105} (2022) 126026} [\href{https://arxiv.org/abs/2112.08433}{{\ttfamily 2112.08433}}].

\bibitem{Omidi:2021opl}
F.~Omidi, \emph{{Entropy of Hawking radiation for two-sided hyperscaling violating black branes}}, \href{https://doi.org/10.1007/JHEP04(2022)022}{\emph{JHEP} {\bfseries 04} (2022) 022} [\href{https://arxiv.org/abs/2112.05890}{{\ttfamily 2112.05890}}].

\bibitem{Cadoni:2021ypx}
M.~Cadoni and A.P.~Sanna, \emph{{Unitarity and Page Curve for Evaporation of 2D AdS Black Holes}},  \href{https://arxiv.org/abs/2106.14738}{{\ttfamily 2106.14738}}.

\bibitem{Hollowood:2021lsw}
T.J.~Hollowood, S.P.~Kumar, A.~Legramandi and N.~Talwar, \emph{{Grey-body factors, irreversibility and multiple island saddles}}, \href{https://doi.org/10.1007/JHEP03(2022)110}{\emph{JHEP} {\bfseries 03} (2022) 110} [\href{https://arxiv.org/abs/2111.02248}{{\ttfamily 2111.02248}}].

\bibitem{Geng:2021hlu}
H.~Geng, A.~Karch, C.~Perez-Pardavila, S.~Raju, L.~Randall, M.~Riojas et~al., \emph{{Inconsistency of islands in theories with long-range gravity}}, \href{https://doi.org/10.1007/JHEP01(2022)182}{\emph{JHEP} {\bfseries 01} (2022) 182} [\href{https://arxiv.org/abs/2107.03390}{{\ttfamily 2107.03390}}].

\bibitem{Yadav:2022fmo}
G.~Yadav, \emph{{Page curves of Reissner\textendash{}Nordstr\"om black hole in HD gravity}}, \href{https://doi.org/10.1140/epjc/s10052-022-10873-1}{\emph{Eur. Phys. J. C} {\bfseries 82} (2022) 904} [\href{https://arxiv.org/abs/2204.11882}{{\ttfamily 2204.11882}}].

\bibitem{Anand:2022mla}
A.~Anand, \emph{{Page curve and island in EGB gravity}}, \href{https://doi.org/10.1016/j.nuclphysb.2023.116284}{\emph{Nucl. Phys. B} {\bfseries 993} (2023) 116284} [\href{https://arxiv.org/abs/2205.13785}{{\ttfamily 2205.13785}}].

\bibitem{Gyongyosi:2022vaf}
Z.~Gyongyosi, T.J.~Hollowood, S.P.~Kumar, A.~Legramandi and N.~Talwar, \emph{{Black Hole Information Recovery in JT Gravity}},  \href{https://arxiv.org/abs/2209.11774}{{\ttfamily 2209.11774}}.

\bibitem{Grimaldi:2022suv}
G.~Grimaldi, J.~Hernandez and R.C.~Myers, \emph{{Quantum extremal islands made easy. Part IV. Massive black holes on the brane}}, \href{https://doi.org/10.1007/JHEP03(2022)136}{\emph{JHEP} {\bfseries 03} (2022) 136} [\href{https://arxiv.org/abs/2202.00679}{{\ttfamily 2202.00679}}].

\bibitem{Du:2022vvg}
D.-H.~Du, W.-C.~Gan, F.-W.~Shu and J.-R.~Sun, \emph{{Unitary constraints on semiclassical Schwarzschild black holes in the presence of island}}, \href{https://doi.org/10.1103/PhysRevD.107.026005}{\emph{Phys. Rev. D} {\bfseries 107} (2023) 026005} [\href{https://arxiv.org/abs/2206.10339}{{\ttfamily 2206.10339}}].

\bibitem{Yu:2022xlh}
M.-H.~Yu and X.-H.~Ge, \emph{{Entanglement islands in generalized two-dimensional dilaton black holes}}, \href{https://doi.org/10.1103/PhysRevD.107.066020}{\emph{Phys. Rev. D} {\bfseries 107} (2023) 066020} [\href{https://arxiv.org/abs/2208.01943}{{\ttfamily 2208.01943}}].

\bibitem{Ageev:2022qxv}
D.S.~Ageev, I.Y.~Aref'eva, A.I.~Belokon, A.V.~Ermakov, V.V.~Pushkarev and T.A.~Rusalev, \emph{{Infrared regularization and finite size dynamics of entanglement entropy in Schwarzschild black hole}}, \href{https://doi.org/10.1103/PhysRevD.108.046005}{\emph{Phys. Rev. D} {\bfseries 108} (2023) 046005} [\href{https://arxiv.org/abs/2209.00036}{{\ttfamily 2209.00036}}].

\bibitem{HosseiniMansoori:2022hok}
S.A.~Hosseini~Mansoori, O.~Luongo, S.~Mancini, M.~Mirjalali, M.~Rafiee and A.~Tavanfar, \emph{{Planar black holes in holographic axion gravity: Islands, Page times, and scrambling times}}, \href{https://doi.org/10.1103/PhysRevD.106.126018}{\emph{Phys. Rev. D} {\bfseries 106} (2022) 126018} [\href{https://arxiv.org/abs/2209.00253}{{\ttfamily 2209.00253}}].

\bibitem{Yadav:2022jib}
G.~Yadav and N.~Joshi, \emph{{Cosmological and black hole islands in multi-event horizon spacetimes}}, \href{https://doi.org/10.1103/PhysRevD.107.026009}{\emph{Phys. Rev. D} {\bfseries 107} (2023) 026009} [\href{https://arxiv.org/abs/2210.00331}{{\ttfamily 2210.00331}}].

\bibitem{Lu:2022tmt}
C.-Y.~Lu, M.-H.~Yu, X.-H.~Ge and L.-J.~Tian, \emph{{Page curve and phase transition in deformed Jackiw\textendash{}Teitelboim gravity}}, \href{https://doi.org/10.1140/epjc/s10052-023-11358-5}{\emph{Eur. Phys. J. C} {\bfseries 83} (2023) 215} [\href{https://arxiv.org/abs/2210.14750}{{\ttfamily 2210.14750}}].

\bibitem{Lu:2022cgq}
Y.~Lu and J.~Lin, \emph{{The Markov gap in the presence of islands}}, \href{https://doi.org/10.1007/JHEP03(2023)043}{\emph{JHEP} {\bfseries 03} (2023) 043} [\href{https://arxiv.org/abs/2211.06886}{{\ttfamily 2211.06886}}].

\bibitem{Basu:2022crn}
D.~Basu, Q.~Wen and S.~Zhou, \emph{{Entanglement Islands from Hilbert Space Reduction}},  \href{https://arxiv.org/abs/2211.17004}{{\ttfamily 2211.17004}}.

\bibitem{Suzuki:2022xwv}
K.~Suzuki and T.~Takayanagi, \emph{{BCFT and Islands in two dimensions}}, \href{https://doi.org/10.1007/JHEP06(2022)095}{\emph{JHEP} {\bfseries 06} (2022) 095} [\href{https://arxiv.org/abs/2202.08462}{{\ttfamily 2202.08462}}].

\bibitem{Karch:2022rvr}
A.~Karch, H.~Sun and C.F.~Uhlemann, \emph{{Double holography in string theory}}, \href{https://doi.org/10.1007/JHEP10(2022)012}{\emph{JHEP} {\bfseries 10} (2022) 012} [\href{https://arxiv.org/abs/2206.11292}{{\ttfamily 2206.11292}}].

\bibitem{Cadoni:2023tse}
M.~Cadoni, M.~Oi and A.P.~Sanna, \emph{{Evaporation and information puzzle for 2D nonsingular asymptotically flat black holes}}, \href{https://doi.org/10.1007/JHEP06(2023)211}{\emph{JHEP} {\bfseries 06} (2023) 211} [\href{https://arxiv.org/abs/2303.05557}{{\ttfamily 2303.05557}}].

\bibitem{Piao:2023vgm}
Y.-S.~Piao, \emph{{Implication of the island rule for inflation and primordial perturbations}}, \href{https://doi.org/10.1103/PhysRevD.107.123509}{\emph{Phys. Rev. D} {\bfseries 107} (2023) 123509} [\href{https://arxiv.org/abs/2301.07403}{{\ttfamily 2301.07403}}].

\bibitem{Guo:2023gfa}
C.-Z.~Guo, W.-C.~Gan and F.-W.~Shu, \emph{{Page curves and entanglement islands for the step-function Vaidya model of evaporating black holes}}, \href{https://doi.org/10.1007/JHEP05(2023)042}{\emph{JHEP} {\bfseries 05} (2023) 042} [\href{https://arxiv.org/abs/2302.02379}{{\ttfamily 2302.02379}}].

\bibitem{Parvizi:2023foz}
S.~Parvizi and M.~Shahbazi, \emph{{Analogue gravity and the island prescription}}, \href{https://doi.org/10.1140/epjc/s10052-023-11874-4}{\emph{Eur. Phys. J. C} {\bfseries 83} (2023) 705} [\href{https://arxiv.org/abs/2302.08742}{{\ttfamily 2302.08742}}].

\bibitem{Hung:2023mbw}
T.N.~Hung and C.H.~Nam, \emph{{Compactified extra dimension and entanglement island as clues to quantum gravity}}, \href{https://doi.org/10.1140/epjc/s10052-023-11606-8}{\emph{Eur. Phys. J. C} {\bfseries 83} (2023) 472} [\href{https://arxiv.org/abs/2303.00348}{{\ttfamily 2303.00348}}].

\bibitem{Jeong:2023hrb}
H.-S.~Jeong, K.-Y.~Kim and Y.-W.~Sun, \emph{{Island in dyonic black holes: doubly holographic theory}},  \href{https://arxiv.org/abs/2305.18122}{{\ttfamily 2305.18122}}.

\bibitem{Tong:2023nvi}
C.-W.~Tong, D.-H.~Du and J.-R.~Sun, \emph{{Island of Reissner-Nordstr$\mathbf{\ddot{o}}$m anti-de Sitter black holes in the large $d$ limit}},  \href{https://arxiv.org/abs/2306.06682}{{\ttfamily 2306.06682}}.

\bibitem{Yu:2023whl}
M.-H.~Yu, X.-H.~Ge and C.-Y.~Lu, \emph{{Page Curves for Accelerating Black Holes}},  \href{https://arxiv.org/abs/2306.11407}{{\ttfamily 2306.11407}}.

\bibitem{Matsuo:2023cmb}
Y.~Matsuo, \emph{{Quantum focusing conjecture and the Page curve}},  \href{https://arxiv.org/abs/2308.05009}{{\ttfamily 2308.05009}}.

\bibitem{Yadav:2022mnv}
G.~Yadav and A.~Misra, \emph{{Entanglement entropy and Page curve from the M-theory dual of thermal QCD above Tc at intermediate coupling}}, \href{https://doi.org/10.1103/PhysRevD.107.106015}{\emph{Phys. Rev. D} {\bfseries 107} (2023) 106015} [\href{https://arxiv.org/abs/2207.04048}{{\ttfamily 2207.04048}}].

\bibitem{Matsuo:2020ypv}
Y.~Matsuo, \emph{{Islands and stretched horizon}}, \href{https://doi.org/10.1007/JHEP07(2021)051}{\emph{JHEP} {\bfseries 07} (2021) 051} [\href{https://arxiv.org/abs/2011.08814}{{\ttfamily 2011.08814}}].

\bibitem{Iizuka:2021tut}
N.~Iizuka, A.~Miyata and T.~Ugajin, \emph{{A comment on a fine-grained description of evaporating black holes with baby universes}}, \href{https://doi.org/10.1007/JHEP09(2022)158}{\emph{JHEP} {\bfseries 09} (2022) 158} [\href{https://arxiv.org/abs/2111.07107}{{\ttfamily 2111.07107}}].

\bibitem{Anegawa:2020ezn}
T.~Anegawa and N.~Iizuka, \emph{{Notes on islands in asymptotically flat 2d dilaton black holes}}, \href{https://doi.org/10.1007/JHEP07(2020)036}{\emph{JHEP} {\bfseries 07} (2020) 036} [\href{https://arxiv.org/abs/2004.01601}{{\ttfamily 2004.01601}}].

\bibitem{Almheiri:2021jwq}
A.~Almheiri and H.W.~Lin, \emph{{The entanglement wedge of unknown couplings}}, \href{https://doi.org/10.1007/JHEP08(2022)062}{\emph{JHEP} {\bfseries 08} (2022) 062} [\href{https://arxiv.org/abs/2111.06298}{{\ttfamily 2111.06298}}].

\bibitem{Afrasiar:2023nir}
M.~Afrasiar, D.~Basu, A.~Chandra, V.~Raj and G.~Sengupta, \emph{{Islands and dynamics at the interface}},  \href{https://arxiv.org/abs/2306.12476}{{\ttfamily 2306.12476}}.

\bibitem{Anand:2023ozw}
A.~Anand, \emph{{Island in Warped AdS Black Holes}},  \href{https://arxiv.org/abs/2308.05432}{{\ttfamily 2308.05432}}.

\bibitem{Kashyap:2023keo}
S.P.~Kashyap, R.~Pius and M.~Ramchander, \emph{{Theory dependence of black hole interior reconstruction and the extended strong subadditivity}},  \href{https://arxiv.org/abs/2306.10801}{{\ttfamily 2306.10801}}.

\bibitem{Blommaert:2023vbz}
A.~Blommaert, J.~Kruthoff and S.~Yao, \emph{{The power of Lorentzian wormholes}}, \href{https://doi.org/10.1007/JHEP10(2023)005}{\emph{JHEP} {\bfseries 10} (2023) 005} [\href{https://arxiv.org/abs/2302.01360}{{\ttfamily 2302.01360}}].

\bibitem{Guo:2023fly}
Y.~Guo and R.-X.~Miao, \emph{{Page curves on codim-m and charged branes}}, \href{https://doi.org/10.1140/epjc/s10052-023-12026-4}{\emph{Eur. Phys. J. C} {\bfseries 83} (2023) 847}.

\bibitem{Chang:2023gkt}
J.-C.~Chang, S.~He, Y.-X.~Liu and L.~Zhao, \emph{{Island formula in Planck brane}},  \href{https://arxiv.org/abs/2308.03645}{{\ttfamily 2308.03645}}.

\bibitem{Li:2023rue}
R.~Li, X.~Wang, K.~Zhang and J.~Wang, \emph{{Retrieving information from Hawking radiation in the non-isometric holographic model of black hole interior: theory and quantum simulations}},  \href{https://arxiv.org/abs/2307.01454}{{\ttfamily 2307.01454}}.

\bibitem{Li:2023fly}
D.~Li and R.-X.~Miao, \emph{{Massless entanglement islands in cone holography}}, \href{https://doi.org/10.1007/JHEP06(2023)056}{\emph{JHEP} {\bfseries 06} (2023) 056} [\href{https://arxiv.org/abs/2303.10958}{{\ttfamily 2303.10958}}].

\bibitem{Akers:2022qdl}
C.~Akers, N.~Engelhardt, D.~Harlow, G.~Penington and S.~Vardhan, \emph{{The black hole interior from non-isometric codes and complexity}},  \href{https://arxiv.org/abs/2207.06536}{{\ttfamily 2207.06536}}.

\bibitem{Murdia:2022giv}
C.~Murdia, Y.~Nomura and K.~Ritchie, \emph{{Black hole and de Sitter microstructures from a semiclassical perspective}}, \href{https://doi.org/10.1103/PhysRevD.107.026016}{\emph{Phys. Rev. D} {\bfseries 107} (2023) 026016} [\href{https://arxiv.org/abs/2207.01625}{{\ttfamily 2207.01625}}].

\bibitem{Chandra:2022fwi}
J.~Chandra and T.~Hartman, \emph{{Coarse graining pure states in AdS/CFT}}, \href{https://doi.org/10.1007/JHEP10(2023)030}{\emph{JHEP} {\bfseries 10} (2023) 030} [\href{https://arxiv.org/abs/2206.03414}{{\ttfamily 2206.03414}}].

\bibitem{Li:2023nfv}
R.~Li, X.~Wang, K.~Zhang and J.~Wang, \emph{{High-fidelity information recovery from radiating black holes through random local projection}},  \href{https://arxiv.org/abs/2309.01917}{{\ttfamily 2309.01917}}.

\bibitem{Hirano:2023ebw}
S.~Hirano, \emph{{Island Formula from Wald-like Entropy with Backreaction}},  \href{https://arxiv.org/abs/2310.03416}{{\ttfamily 2310.03416}}.

\bibitem{Hirano:2021rzg}
S.~Hirano and T.~Kuroki, \emph{{Replica wormholes from Liouville theory}}, \href{https://doi.org/10.1007/JHEP01(2022)094}{\emph{JHEP} {\bfseries 01} (2022) 094} [\href{https://arxiv.org/abs/2109.12539}{{\ttfamily 2109.12539}}].

\bibitem{Okuyama:2021bqg}
K.~Okuyama and K.~Sakai, \emph{{Page curve from dynamical branes in JT gravity}}, \href{https://doi.org/10.1007/JHEP02(2022)087}{\emph{JHEP} {\bfseries 02} (2022) 087} [\href{https://arxiv.org/abs/2111.09551}{{\ttfamily 2111.09551}}].

\bibitem{Chou:2021boq}
C.-J.~Chou, H.B.~Lao and Y.~Yang, \emph{{Page curve of effective Hawking radiation}}, \href{https://doi.org/10.1103/PhysRevD.106.066008}{\emph{Phys. Rev. D} {\bfseries 106} (2022) 066008} [\href{https://arxiv.org/abs/2111.14551}{{\ttfamily 2111.14551}}].

\bibitem{Chou:2023adi}
C.-J.~Chou, H.B.~Lao and Y.~Yang, \emph{{Page Curve of AdS-Vaidya Model for Evaporating Black Holes}},  \href{https://arxiv.org/abs/2306.16744}{{\ttfamily 2306.16744}}.

\bibitem{Geng:2022slq}
H.~Geng, A.~Karch, C.~Perez-Pardavila, S.~Raju, L.~Randall, M.~Riojas et~al., \emph{{Jackiw-Teitelboim Gravity from the Karch-Randall Braneworld}}, \href{https://doi.org/10.1103/PhysRevLett.129.231601}{\emph{Phys. Rev. Lett.} {\bfseries 129} (2022) 231601} [\href{https://arxiv.org/abs/2206.04695}{{\ttfamily 2206.04695}}].

\bibitem{Gyongyosi:2023sue}
Z.~Gyongyosi, T.J.~Hollowood, S.P.~Kumar, A.~Legramandi and N.~Talwar, \emph{{The holographic map of an evaporating black hole}}, \href{https://doi.org/10.1007/JHEP07(2023)043}{\emph{JHEP} {\bfseries 07} (2023) 043} [\href{https://arxiv.org/abs/2301.08362}{{\ttfamily 2301.08362}}].

\bibitem{Basu:2023wmv}
D.~Basu, J.~Lin, Y.~Lu and Q.~Wen, \emph{{Ownerless island and partial entanglement entropy in island phases}},  \href{https://arxiv.org/abs/2305.04259}{{\ttfamily 2305.04259}}.

\bibitem{Karch:2002sh}
A.~Karch and E.~Katz, \emph{{Adding flavor to AdS / CFT}}, \href{https://doi.org/10.1088/1126-6708/2002/06/043}{\emph{JHEP} {\bfseries 06} (2002) 043} [\href{https://arxiv.org/abs/hep-th/0205236}{{\ttfamily hep-th/0205236}}].

\bibitem{Chakrabortty:2011sp}
S.~Chakrabortty, \emph{{Dissipative force on an external quark in heavy quark cloud}}, \href{https://doi.org/10.1016/j.physletb.2011.09.112}{\emph{Phys. Lett. B} {\bfseries 705} (2011) 244} [\href{https://arxiv.org/abs/1108.0165}{{\ttfamily 1108.0165}}].

\bibitem{Chakrabortty:2016xcb}
S.~Chakrabortty and T.K.~Dey, \emph{{Back reaction effects on the dynamics of heavy probes in heavy quark cloud}}, \href{https://doi.org/10.1007/JHEP05(2016)094}{\emph{JHEP} {\bfseries 05} (2016) 094} [\href{https://arxiv.org/abs/1602.04761}{{\ttfamily 1602.04761}}].

\bibitem{Chakrabortty:2020ptb}
S.~Chakrabortty, S.~Pant and K.~Sil, \emph{{Effect of back reaction on entanglement and subregion volume complexity in strongly coupled plasma}}, \href{https://doi.org/10.1007/JHEP06(2020)061}{\emph{JHEP} {\bfseries 06} (2020) 061} [\href{https://arxiv.org/abs/2004.06991}{{\ttfamily 2004.06991}}].

\bibitem{Chakrabortty:2022kvq}
S.~Chakrabortty, H.~Hoshino, S.~Pant and K.~Sil, \emph{{A holographic study of the characteristics of chaos and correlation in the presence of backreaction}}, \href{https://doi.org/10.1016/j.physletb.2023.137749}{\emph{Phys. Lett. B} {\bfseries 838} (2023) 137749} [\href{https://arxiv.org/abs/2206.12555}{{\ttfamily 2206.12555}}].

\bibitem{Hayden:2007cs}
P.~Hayden and J.~Preskill, \emph{{Black holes as mirrors: Quantum information in random subsystems}}, \href{https://doi.org/10.1088/1126-6708/2007/09/120}{\emph{JHEP} {\bfseries 09} (2007) 120} [\href{https://arxiv.org/abs/0708.4025}{{\ttfamily 0708.4025}}].

\bibitem{Sekino:2008he}
Y.~Sekino and L.~Susskind, \emph{{Fast Scramblers}}, \href{https://doi.org/10.1088/1126-6708/2008/10/065}{\emph{JHEP} {\bfseries 10} (2008) 065} [\href{https://arxiv.org/abs/0808.2096}{{\ttfamily 0808.2096}}].

\end{thebibliography}\endgroup

\end{document}